\documentstyle[12pt]{article}

\def\half{{\textstyle\frac{1}{2}}}
\def\ihalf{{\textstyle\frac{i}{2}}}
\def\re{{\rm Re}}

\def\Q{Q}

\def\T{T}
\def\R{R}
\def\S{S}
\newcommand{\sgn}[1]{[#1]}
\newcommand{\ve}[1]{\hbox{\boldmath{$#1$}}}
\newcommand{\sve}[1]{\hbox{\boldmath{$\scriptstyle #1$}}}
\newcommand{\ma}[1]{\hbox{\boldmath{$\rm #1$}}}
\newcommand{\Op}[1]{\hat{#1}}
\newcommand{\Oo}[1]{\acute{#1}}
\newcommand{\oO}[1]{\grave{#1}}
\newcommand{\op}[1]{\check{#1}}

\newcommand{\Bra}[1]{\langle #1|}
\newcommand{\Ket}[1]{|#1\rangle}
\newcommand{\Braket}[2]{\langle #1|#2\rangle}
\newcommand{\bra}[1]{\{ #1|}
\newcommand{\ket}[1]{|#1\} }
\newcommand{\braket}[2]{\{ #1|#2\} }

\begin{document}
\begin{flushright}
Preprint CAMTP/94-9\\
November 1994\\
\end{flushright}
\begin{center}
\vspace{1in}
\Large
{\bf General Quantum Surface of Section Method}\\
\vspace{0.5in}
\large
Toma\v z Prosen\footnote{e-mail: Tomaz.Prosen@UNI-MB.SI}\\
\normalsize
\vspace{0.3in}
Center for Applied Mathematics and Theoretical Physics,\\
University of Maribor, Krekova 2, SLO-62000 Maribor, Slovenia\\
\end{center}
\vspace{0.5in}

\noindent{\bf Abstract}
A new method for {\em exact} quantization of general
bound Hamiltonian systems is presented. It is the quantum analogue of the
classical Poincar\' e
Surface Of Section (SOS) reduction of classical dynamics. The quantum
Poincar\' e mapping is shown to be the product of the two
{\em generalized} ({\em non-unitary} but {\em compact})
on-shell scattering operators
of the two scattering Hamiltonians which are obtained from the original bound
one by cutting the $f$-dim configuration space (CS) along $(f-1)-$dim
configurational SOS and attaching the flat quasi-one-dimensional waveguides
instead. The quantum Poincar\' e mapping has fixed points at the eigenenergies
of the original bound Hamiltonian. The energy dependent quantum propagator
$(E-\Op{H})^{-1}$ can be decomposed in terms of the four energy dependent
propagators which propagate from and/or to CS to and/or
from configurational SOS (which may generally be composed of many
disconnected parts).

I show that in the semiclassical limit ($\hbar\rightarrow 0$)
the quantum Poincar\' e mapping converges to the Bogomolny's propagator
and explain how the higher order semiclassical corrections can be
obtained systematically.


\bigskip

\noindent Submitted to {\bf Journal of Physics A: Mathematical and General}
\newpage

\section{Introduction}

Over the last decade or two there has been an increasing interest in
efficient quantization procedures for simple (having only few freedoms)
but nonlinear (possibly chaotic) Hamiltonian systems. Here I consider
{\em bound} and {\em autonomous} Hamiltonian systems with $f$ freedoms.
Directly solving the time-independent Schr\" odinger equation in $f-$dim
configuration space (CS) or the equivalent eigenvalue problem
for the Hamiltonian matrix in an appropriate basis is the first but certainly
not the best idea. The question appeared \cite{A94} whether there exists
a quantum analogue of surface of section (SOS) reduction of classical dynamics
\cite{LL83} which reduces smooth bound and autonomous Hamiltonian
dynamics over $2f$-dim phase space to a discrete Poincar\' e mapping over
only $(2f-2)$-dim SOS.

In the case of quantum billiard systems in two dimensions ($f=2$) we have
the so called boundary integral method which reduces 2-dim
Schr\" odinger equation to 1-dim integral equation. Its kernel can
be interpreted as the quantum bounce map which is a special case of
Poincar\' e mapping. Smilansky and coworkers
\cite{DS92,SS93} have developed more general {\em scattering approach
for quantization of billiards}. They construct exact quantum Poincar\' e
mapping for 2-dim billiards with respect to the arbitrary line of section
as the product of the two scattering matrices of the two opened billiards.
These methods are typically much more efficient than the direct
diagonalization,
since the dimension of the matrices that they use is typically of the order of
square root of the dimension of the original Hamiltonian matrix.
On the other hand Bogomolny succeeded to construct approximate semiclassical
Poincar\' e mapping with respect to arbitrary configurational surface
of section for arbitrary autonomous Hamiltonian.
In this paper I present the generalization of the scattering approach
for quantization of almost arbitrary bound Hamiltonians and show that it
reduces to the Bogomolny's theory in the semiclassical limit
$\hbar\rightarrow 0$.

In section 2 I construct the quantum Poincar\' e
mapping and prove that the eigenenergies of the original Hamiltonian
correspond to the fixed points of quantum Poincar\' e mapping. I prove also
theoretically perhaps even more interesting SOS decomposition of the resolvent
of the Hamiltonian.  Here I also study the semiclassical limit of newly
defined propagators and explicitly calculate the leading order and
next-to-leading order terms while I explain how higher order corrections (in
powers of $\hbar$) can be systematically obtained.
The symmetries of quantum Poincar\' e mapping are discussed and
it is explained how the SOS quantization condition can be very efficiently
used in practical calculations, especially for the generic class of the
so called {\em semi-separable systems}.
In section 3 I formulate an abstract quantum SOS
method which can be applied to {\em arbitrary boundary value
differential equation} problems. Then I apply an abstract theory to the
case of energy-dependent Schr\" odinger equation of section 2 and
more general cases of nonrelativistic or even relativistic systems (described
by
the Dirac equation) coupled to arbitrary external gauge fields. In section 4
the method is generalized to the case of non-simply but multiply connected
CSOS. In section 5 I discuss the meaning and applicability of the new results
and conclude.

Some preliminary results of this
project have already been reported \cite{P94a,P94b}. Recently I have been
informed that one of the results of this paper, namely the SOS
quantization condition for 2-dim Hamiltonian systems of the standard type,
has also been obtained independently and subsequently by other authors
\cite{RS94}.

\section{Surface of Section Quantization}

\subsection{Notation}

The basic results of this paper are most beautifully and compactly written
in terms of some new physical quantities whose mathematical definitions and
notation are described in this subsection.

We study {\em autonomous} and
{\em bound} (at least in the energy region of our concern) Hamiltonian
systems with few, say $f$, freedoms, living in an $f-$dim {\em configuration
space (CS) ${\cal C}$. One should also provide a smooth $(f-1)$-dim
submanifold} of CS ${\cal C}$ which shall be called {\em configurational
surface of section} (CSOS)
\footnote{The more general case of $(2f-2)$-dim SOS in $2f$-dim phase
space which is not perpendicular to CS cannot be treated within the
present approach except in the cases where one can change the
phase space coordinates by means of an appropriate canonical transformation.}
and denoted by ${\cal S}_0$. In this section
we consider only the case of {\em simply-connected} CSOS whereas in
section 4 we study the case of more general {\em multiply-connected} CSOS.
We choose the coordinates in CS, $\ve{q} = (\ve{x},y) \in {\cal C}$ in such
a way that the CSOS is given by a simple constraint $y=0$, or ${\cal S}_0 =
({\cal S},0).$ These coordinates {\em need not be global}, i.e. they need not
cover the whole CS, but they should cover the open set which
includes the whole CSOS ${\cal S}_0$. This means that every point in
${\cal S}_0$ should be uniquely represented by
CSOS coordinates $\ve{x}\in {\cal S}$ which {\em may be more general} than
Euclidean coordinates ${\cal R}^{f-1}$ (e.g. $(f-1)$ dim sphere $\S^{f-1}$).
In this section we shall assume that ${\cal S}_0$ is an {\em orientable
manifold} so that it cuts the CS ${\cal C}$ in two pieces which will be
referred to as upper and lower and denoted by the value of the binary
index $\sigma = \uparrow,\downarrow$ (see figure 1). In arithmetic expressions
the arrows will have the following values $\uparrow=+1,\downarrow=-1$.
My approach presented in this section applies to quite general class
of bound Hamiltonians whose kinetic energy is quadratic at least
perpendicularly to CSOS
\begin{equation}
H = \frac{1}{2m} p_y^2 + H^\prime (\ve{p}_x,\ve{x},y).
\label{eq:clH}
\end{equation}
In next sections we generalize this class to include Hamiltonians
having coordinate-dependent mass (which arise in the {\em curvilinear
coordinates which must be used in case of non-flat CSOS}) and/or terms linear
in $p_y$ (which appear e.g. due to the {\em presence of magnetic field}).

In quantum mechanics, the observables are represented by self-adjoint
operators in a Hilbert space ${\cal H}$ of complex-valued functions
$\Psi(\ve{q})$ over the CS ${\cal C}$ which obey
boundary conditions $\Psi(\partial {\cal C})=0$
\footnote{In case of infinite CS (e.g. Euclidean space ${\cal C}=\R^f$)
we assume that points at infinity belong to the boundary $\partial{\cal C}$.}
and have finite $L^2$ norm $\int_{\cal C} d\ve{q}|\Psi(\ve{q})|^2 < \infty$.
We shall use the Dirac's notation. Pure state of a
physical system is represented by a vector --- {\em ket} $\Ket{\Psi}$
which can be expanded in a convenient complete set of basis vectors, e.g.
position eigenvectors $\Ket{\ve{q}}=\Ket{\ve{x},y},\;
\Ket{\Psi} = \int_{\cal C}d\ve{q}\Ket{\ve{q}}\Braket{\ve{q}}{\Psi}=
\int_{\cal C}d\ve{q}\Psi(\ve{q})\Ket{\ve{q}}$
(in a symbolic sense, since $\Ket{\ve{q}}$ are not proper vectors,
but such expansions are still meaningful iff
$\Psi(\ve{q})=\Braket{\ve{q}}{\Psi}$ is square integrable i.e.
$L^2({\cal C})$-function). Every ket $\Ket{\Psi}\in{\cal H}$ has a
corresponding
vector from the dual Hilbert space ${\cal H}^\prime$, that is {\em bra}
$\Bra{\Psi}\in{\cal H}^\prime,\; \Braket{\Psi}{\ve{q}} =
\Braket{\ve{q}}{\Psi}^*$. We shall use mathematical accent $\Op{}$ to denote
linear operators over the Hilbert space ${\cal H}$.
Operators of SOS coordinates $\Op{\ve{x}}$ and $\Op{\ve{p}}_{\ve{x}}$, defined
by\footnote{In case of non-euclidean SOS they should be replaced by the
generators of the corresponding Lie algebra.}
\begin{eqnarray*}
\Bra{\ve{x},y}\Op{\ve{x}}\Ket{\Psi} &=& \ve{x}\Psi(\ve{x},y),\\
\Bra{\ve{x},y}\Op{\ve{p}}_{\ve{x}}\Ket{\Psi} &=& -i\hbar\partial_{\ve{x}}
\Psi(\ve{x},y),
\end{eqnarray*}
can be viewed also as acting on functions $\psi(\ve{x})$ of $\ve{x}$ only
and therefore operating in some other, much smaller Hilbert space of
square-integrable complex-valued functions over a
CSOS ${\cal S}_0$
\begin{eqnarray*}
\bra{\ve{x}}\op{\ve{x}}\ket{\psi} &=& \ve{x}\psi(\ve{x}),\label{eq:xSOS}\\
\bra{\ve{x}}\op{\ve{p}}_{\ve{x}}\ket{\psi} &=& -i\hbar\partial_{\ve{x}}
\psi(\ve{x}). \label{eq:pSOS}
\end{eqnarray*}
Vectors in such SOS-Hilbert space, denoted by ${\cal L}$,
will be written as $\ket{\psi}$ and linear operators over ${\cal L}$
will wear mathematical accent $\op{}$ like
restricted position $\op{\ve{x}}$ and momentum $\op{\ve{p}}_{\ve{x}}$.
Eigenvectors $\ket{\ve{x}}$ of SOS-position operator $\op{\ve{x}}$
provide a useful complete set of basis vectors of ${\cal L}$.
The quantum Hamiltonian can be written as
\begin{equation}
\Op{H} = -\frac{\hbar^2}{2m}\partial_y^2 + \Op{H}^\prime(y),\quad
\Op{H}^\prime(y) = H^\prime(-i\hbar\partial_{\ve{x}},\ve{x},y).
\label{eq:qH}
\end{equation}
The eigenstates of the {\em inside-CSOS Hamiltonian}
$\op{H}^\prime(0) = \Op{H}^\prime(0)\vert_{\cal L}$
restricted to the SOS-Hilbert space ${\cal L}$,
$\ket{n}\in{\cal L}$
\begin{equation}
\op{H}^\prime(0)\ket{n} = E^\prime_n\ket{n},
\label{eq:eigenmodes}
\end{equation}
which are called {\em SOS-eigenmodes},
provide a useful (countable $n=1,2,\ldots$) complete and orthogonal basis
for ${\cal L}$ since $\op{H}^\prime(0)$ is a self-adjoint
operator with discrete spectrum when its domain is restricted to ${\cal L}$.

The major problem of bound quantum dynamics is to determine the
{\em eigenenergies} $E$ for which the {\em Schr\" odinger equation}
\begin{equation}
\Bra{\ve{x},y}\Op{H}\Ket{\Psi_\sigma(E)} = E\Psi_\sigma(\ve{x},y,E)
\label{eq:Schreq}
\end{equation}
has nontrivial normalizable solutions --- {\em eigenfunctions}
$\Psi(\ve{x},y,E)$.

\subsection{Scattering formulation}

In this subsection I will introduce our basic tools using the powerful
quantum mechanical time-independent multi-channel scattering theory
\cite{N82,W64}.

To connect bound Hamiltonian dynamics and scattering theory one should make the
following very important step. Cut one part of CS off along CSOS and attach
a semi-infinite separable (flat along the $y-axis$) waveguide instead
(see figure 1). Thus we introduce two scattering Hamiltonians
\begin{equation}
\Op{H}_\sigma = \left\{ \begin{array}{ll}
-(\hbar^2/2m)\partial_y^2 + \Op{H}^\prime(y); & \sigma y \ge 0,\\
-(\hbar^2/2m)\partial_y^2 + \Op{H}^\prime(0); & \sigma y < 0.\end{array}
\right.
\label{eq:scH}
\end{equation}
Every wavefunction inside the waveguide ($\sigma y \le 0$)
at energy $E$ can be separated as the superposition of
products of a bound state (SOS-eigenmode $n$) in $\ve{x}$
direction and free motion in $y-$direction,
$$ \braket{x}{n} e^{\pm i k_n(E) y}, $$
with the corresponding wavenumber determined by the energy difference
$E-E^\prime_n$ available for the motion perpendicular to the CSOS
$$ k_n(E) = \sqrt{\frac{2m}{\hbar^2} (E - E^\prime_n)}. $$
For any value of energy $E$, there is typically a finite number of the
so called open or propagating SOS-eigenmodes --- channels with real
wavenumbers for which $E_n^\prime < E$, and an infinitely many closed channels
with imaginary wavenumbers for which $E_n^\prime > E$. Scattering wavefunction
$\Psi_{\sigma}(\ve{x},y,E)$ at a given energy $E$ (or complex conjugated
wavefunction $\Psi^*_\sigma(\ve{x},y,E^*) = (\Psi_\sigma(\ve{x},y,E^*))^*$)
satisfying the Schr\" odinger equation
$\Op{H}_\sigma\Ket{\Psi_\sigma(E)} = E\Ket{\Psi_\sigma(E)}$
can be uniquely parametrized by vector $\ket{\psi}$ from the SOS-Hilbert
space ${\cal L}$ (or by vector $\bra{\psi^*}$ from the dual SOS-Hilbert
space ${\cal L}^\prime$). $\ket{\psi}\in{\cal L}$ essentially parametrize
the incoming waves
$$\Psi^{\rm in}_\sigma(\ve{x},y,E) = \sum_n \braket{\ve{x}}{n}
\sqrt{\frac{m}{i\hbar^2 k_n(E)}} e^{i\sigma k_n(E)y}\braket{n}{\psi} $$
which uniquely determine the whole scattering wavefunction.
Therefore the {\em wave operators} can be defined, namely
$\Oo{Q}^\prime_\sigma(E)$ which map from
${\cal L}$ to ${\cal H}$
(or $\oO{P}^\prime_\sigma(E)$ which map from
${\cal H}$ to ${\cal L}$) and whose kernels are given
by the scattering wavefunctions (or their complex conjugates)
\begin{eqnarray}
\Bra{\ve{q}}\Oo{Q}^\prime_\sigma(E)\ket{\psi} &=& \Psi_\sigma(\ve{q},E),\\
\bra{\psi^*}\oO{P}^\prime_\sigma(E)\Ket{\ve{q}} &=& \Psi^*_\sigma(\ve{q},E^*).
\label{eq:defPp}
\end{eqnarray}
On the $\sigma-$side of CS ($\sigma y \ge 0$) the scattering wavefunction
{\em satisfies ordinary Schr\" o\-din\-ger equation} (\ref{eq:Schreq})
whereas in the waveguide ($\sigma y \le 0$) it is a
superposition of incoming and scattered waves
\begin{eqnarray}
\Psi_\sigma(\ve{x},y,E) &=& \frac{\sqrt{-i m}}{\hbar}\sum\limits_{n,l}
\braket{\ve{x}}{n}k_n^{-1/2}(E)\left[e^{i\sigma k_n(E)y}\delta_{nl}
+ e^{-i\sigma k_n(E)y}T^\sigma_{nl}\right]\braket{l}{\psi} = \nonumber \\
&=& \frac{\sqrt{-i m}}{\hbar}\bra{\ve{x}}\op{K}^{-1/2}(E)\left[
e^{i\sigma\op{K}(E) y} +
e^{-i\sigma\op{K}(E) y}\op{T}_\sigma(E)\right]\ket{\psi}, \label{eq:scWF}\\
\Psi^*_\sigma(\ve{x},y,E^*) &=&
\frac{\sqrt{-i m}}{\hbar}\bra{\psi^*}\left[
e^{i\sigma\op{K}(E) y} + \op{T}_\sigma(E)e^{-i\sigma\op{K}(E) y}\right]
\op{K}^{-1/2}(E)\ket{\ve{x}}.
\label{eq:scWFc}
\end{eqnarray}
For the sake of compact notation we have introduced the {\em wavenumber
operator}
\begin{equation}
\op{K}(E) = \sum\limits_n k_n(E) \ket{n}\bra{n} =
\sqrt{\frac{2m}{\hbar^2}(E - \op{H}^\prime(0))}.
\end{equation}
$T^\sigma_{nl}(E)$ is the generalized {\em scattering matrix} since it
includes also closed --- non-propagating modes and $\op{T}_\sigma(E)$ is the
corresponding scattering operator over ${\cal L}$
\begin{equation}
\op{T}_\sigma(E) = \sum\limits_{n,l} T^\sigma_{nl} \ket{n}\bra{l}.
\end{equation}
Here I have to make three important notes:
\begin{itemize}
\item Conjugated energy $E^*$ is used in argument of complex
conjugated wavefunction (\ref{eq:defPp}) in order to make all the relevant
operators, e.g. $\oO{P}^\prime_\sigma(E)$, complex analytic functions of $E$
rather than $E^*$.
\item The SOS-states $\ket{\psi^*}$ and $\ket{\psi}$ are generally different.
\item The equation (\ref{eq:scWFc}) is non-trivial and it does not follow from
eq. (\ref{eq:scWF}) but it is a consequence of the {\em Hermitian} symmetry of
the scattering Hamiltonians $\Op{H}_\sigma$ as will be shown in the next
paragraph.
\end{itemize}

Let us now consider the resolvents of the scattering Hamiltonians
(\ref{eq:scH})
with outgoing boundary conditions
\begin{equation}
\Op{G}_\sigma(E) = \frac{1}{i\hbar}\int_0^\infty dt
e^{i(E-\Op{H}_\sigma)t/\hbar} = (E - \Op{H}_\sigma + i 0)^{-1}.
\end{equation}
It is convenient to introduce a hybrid representation of these
scattering Green functions denoted by $\op{G}_\sigma(y,y^\prime,E)\in{\cal L}$
(being a matrix element in $y-$variable and an operator in
$\ve{x}$-variable) defined as
\begin{equation}
\bra{\ve{x}}\op{G}_\sigma(y,y^\prime,E)\ket{\ve{x}^\prime} =
\Bra{\ve{x},y}\Op{G}_\sigma(E)\Ket{\ve{x}^\prime,y^\prime}.
\label{eq:hybG}
\end{equation}
Inside the waveguide ($\sigma y \le 0,\sigma y^\prime \le 0$)
these hybrid Green functions satisfy the following ``free motion''
Schr\" odinger equations in both arguments
\begin{eqnarray}
\partial^2_y \op{G}_\sigma(y,y^\prime,E) + \op{K}^2(E)
\op{G}_\sigma(y,y^\prime,E)&=& \frac{2m}{\hbar^2}\delta(y-y^\prime),
\label{eq:freeSchr}\\
\partial^2_{y^\prime} \op{G}_\sigma(y,y^\prime,E) + \op{G}_\sigma(y,y^\prime,E)
\op{K}^2(E) &=& \frac{2m}{\hbar^2}\delta(y-y^\prime).
\label{eq:freeSchrc}
\end{eqnarray}
The general solution of this linear system (in the waveguide)
is given by the sum of {\em particular} --- ``free motion'' solution
$$\op{G}_{\rm free}(y,y^\prime,E) = \frac{m}{i\hbar^2}
\op{K}^{-1/2}(E)e^{i\op{K}(E)|y-y^\prime|}\op{K}^{-1/2}(E)$$
and general solution of the homogeneous system satisfying outgoing boundary
conditions
\begin{equation}
\op{G}_\sigma(y,y^\prime,E) - \op{G}_{\rm free}(y,y^\prime,E) =
\frac{m}{i\hbar^2}
\op{K}^{-1/2}(E)e^{-i\sigma\op{K}(E)y}\op{A}e^{-i\sigma\op{K}(E)y^\prime}
\op{K}^{-1/2}(E).
\label{eq:Gdiff}
\end{equation}
Their sum $\bra{\ve{x}}\op{G}_\sigma(y,0,E)$ satisfies the Schr\" odinger
equation (\ref{eq:Schreq}) on the $\sigma-$side ($\sigma y \ge 0$), so
comparing it locally, at $\sigma y = +0$, with wavefunctions (\ref{eq:scWF})
yields
\begin{equation}
\Psi_\sigma(\ve{x},y,E) =
\frac{\hbar}{\sqrt{-im}}\bra{\ve{x}}\op{G}_\sigma(y,0,E)\op{K}^{1/2}(E)
\ket{\psi},\quad \sigma y \ge 0,
\label{eq:WFG}
\end{equation}
and determines the free operator valued parameter, $\op{A} = \op{T}_\sigma(E)$.
Thus the waveguide expression for the hybrid scattering Green function reads
($\sigma y \le 0, \sigma y^\prime \le 0$)
\begin{equation}
\op{G}_\sigma(y,y^\prime,E) =
\frac{m}{i\hbar^2}\op{K}^{-1/2}(E)\left[e^{i\op{K}(E)|y-y^\prime|}
+ e^{-i\sigma\op{K}(E)y}\op{T}_\sigma(E)e^{-i\sigma\op{K}(E)y^\prime}\right]
\op{K}^{-1/2}(E).
\label{eq:G}
\end{equation}
Since $\op{G}_\sigma(0,y,E)\ket{\ve{x}}$ satisfies conjugated Schr\" odinger
equation, there exist SOS-states $\ket{\psi^*}$ such that
\begin{equation}
\Psi^*_\sigma(\ve{x},y,E^*) =
\frac{\hbar}{\sqrt{-im}}\bra{\psi^*}\op{K}^{1/2}(E)\op{G}_\sigma(0,y,E)
\ket{\ve{x}},
\quad \sigma y \ge 0,
\label{eq:WFGc}
\end{equation}
and equation (\ref{eq:scWFc}) follows.

\subsection{SOS energy quantization}

Now I shall formulate an exact energy quantization
condition for the original Hamiltonian matrix $\op{H}$ solely in terms
of the scattering operators $\op{T}_\sigma(E)$.
\\\\
{\bf Theorem 1a:}
{\em Every energy $E$ for which the operator
$1 - \op{T}_\downarrow(E)\op{T}_\uparrow(E)$ (where the order of the
arrows may be reversed) is singular is
\begin{itemize}
\item either eigenenergy of the original Hamiltonian $\Op{H}$,
\item or it is a threshold energy for opening of a new channel,
\item or both.
\end{itemize}
More precisely:
The dimensions of the left and right null-space of an
operator $1 - \op{T}_\downarrow(E)\op{T}_\uparrow(E)$ are the same
\begin{equation}
d_T(E) = \dim\ker(1 - \op{T}_\downarrow(E)\op{T}_\uparrow(E)) =
\dim\ker(1 - \op{T}_\downarrow(E)\op{T}_\uparrow(E))^\dagger
\end{equation}
and the following inequality for $d_T(E)$ in terms of dimension of the null
space of operator $E - \Op{H},\;d_H(E) = \dim\ker(E - \Op{H})$, and dimension
of the null space of operator $\op{K}^2(E),\;d_K(E) = \dim\ker\op{K}^2(E)$,
holds
\begin{equation}
\max\{d_H(E),d_K(E)\} \le d_T(E) \le d_H(E) + d_K(E). \label{eq:ineq}
\end{equation}}
\\\\
{\bf Proof:}
Let $d_T(E_0)$ SOS-states $\ket{\uparrow n}\in {\cal L},\, n=1,\ldots,d_T(E_0)$
span the null space of $1 - \op{T}_\downarrow(E_0)\op{T}_\uparrow(E_0)$
\begin{equation}
\left(1 - \op{\T}_\downarrow(E_0)\op{\T}_\uparrow(E_0)\right)
\ket{\uparrow n} = 0. \label{eq:qc2}
\end{equation}
Then one may define another set of $d_T(E_0)$ SOS-states $\ket{\downarrow n}\in
{\cal L}$ by the prescription
\begin{equation}
\ket{\downarrow n} = \op{\T}_\uparrow(E_0) \ket{\uparrow n},
\label{eq:applT1}
\end{equation}
in terms of which the equation (\ref{eq:qc2}) may be rewritten as
a relation symmetric to (\ref{eq:applT1})
\begin{equation}
\ket{\uparrow n} = \op{\T}_{\downarrow}(E_0) \ket{\downarrow n}.
\label{eq:applT2}
\end{equation}
Each of these vectors $\ket{\uparrow n}$ lies either in the null space or in
the image of $\op{K}^2(E_0)$, since
$$ {\cal L} = \ker\op{K}^2(E_0) \oplus \op{K}^2(E_0){\cal L}, $$
where $\op{K}^2(E){\cal L} = \ker\op{K}^2(E)^\perp$ since $\op{K}^2(E)$
is self-adjoint. Let the first $m_K$ vectors
$\ket{\uparrow m},\,m =1,\ldots,m_K$ lie in $\ker\op{K}^2(E)$.
In order to make sure that scattering wavefunctions
(\ref{eq:scWF},\ref{eq:scWFc}) have regular limit $E\rightarrow E_0$, since
$\op{K}^{-1/2}(E)$ is becoming singular if $\ker\op{K}^2(E_0)\ne\emptyset$,
one should demand
\begin{equation}
(1 + \op{T}_\sigma(E_0))\ket{\phi} = 0, \quad
\bra{\phi}(1 + \op{T}_\sigma(E_0)) = 0, \label{eq:aaa}
\end{equation}
for any $\ket{\phi}\in\ker\op{K}^2(E_0)$, so $\ker\op{K}^2(E_0)$ is invariant
under $\op{T}_\sigma(E_0)$ and $\op{T}^\dagger_\sigma(E_0)$
$$
\op{T}_\sigma(E_0) \ker\op{K}^2(E_0) =
\op{T}^\dagger_\sigma(E_0) \ker\op{K}^2(E_0) = \ker\op{K}^2(E_0).
$$
{}From (\ref{eq:aaa}) we also see that
$\ker\op{K}^2(E_0) \subseteq\ker(1 -
\op{\T}_{\downarrow}(E_0)\op{\T}_\uparrow(E_0))$,
so $\ket{\uparrow m}$ span the entire space $\ker\op{K}^2(E_0)$, and so
$d_K(E_0) = m_K \le d_T(E_0)$. Therefore the image $\op{K}^2(E){\cal L}$ is
also invariant under $\op{T}_\sigma(E_0)$ and $\op{T}^\dagger_\sigma(E_0)$,
so the counterparts $\ket{\downarrow l}$ of the remaining
$d_T(E_0)-d_K(E_0)$ SOS-states
$\ket{\uparrow l},\, l=d_K(E_0)+1,\ldots,d_T(E_0)$ from the image
$\op{K}^2(E){\cal L}$ also lie in the image $\op{K}^2(E){\cal L}$.
In the image one can define the inverse of $\op{K}^2(E_0)$ and the
inverse of its fourth root, namely $\op{K}^{-1/2}(E_0)$.
Using (\ref{eq:applT1},\ref{eq:applT2}) one can write
\begin{eqnarray*}
\left(1 + \op{\T}_\uparrow(E_0)\right)\ket{\uparrow l} &=&
\left(1 + \op{\T}_\downarrow(E_0)\right)\ket{\downarrow l}, \label{eq:cr1}\\
\left(1 - \op{\T}_\uparrow(E_0)\right)\ket{\uparrow l} &=&
-\left(1 - \op{\T}_\downarrow(E_0)\right)\ket{\downarrow l}.
\end{eqnarray*}
which can be rewritten using the values of the wavefunctions and
their normal derivatives on the CSOS (from (\ref{eq:scWF}))
\begin{eqnarray}
\Bra{\ve{x},0}\Oo{\Q}^\prime_\sigma(E)\ket{\psi} &=&
\frac{\sqrt{-im}}{\hbar}\bra{\ve{x}}\op{K}^{-1/2}(E)\left(1 +
\op{\T}_\sigma(E)\right)\ket{\psi},
\label{eq:valQ}\\
\partial_y \Bra{\ve{x},y}\Oo{\Q}^\prime_\sigma(E)\ket{\psi}
\vert_{y = 0} &=&
\sigma\frac{\sqrt{im}}{\hbar}\bra{\ve{x}}\op{K}^{1/2}(E)\left(1 -
\op{\T}_\sigma(E)\right)\ket{\psi}
\label{eq:derQ}
\end{eqnarray}
as the continuity of the wavefunctions and their normal derivatives
on the CSOS
\begin{eqnarray*}
\Bra{\ve{x},0}\Oo{\Q}^\prime_\uparrow(E_0)\ket{\uparrow l} &=&
\Bra{\ve{x},0}\Oo{\Q}^\prime_\downarrow(E_0)\ket{\downarrow l} \\
\partial_y \Bra{\ve{x},y}\Oo{\Q}^\prime_\uparrow(E_0)\ket{\uparrow l}
\vert_{y=0} &=&
\partial_y \Bra{\ve{x},y}\Oo{\Q}^\prime_\downarrow(E_0)\ket{\downarrow l}
\vert_{y=0}.
\end{eqnarray*}
which are built up from the pairs of scattering wavefunctions
\begin{equation}
\Psi_n(\ve{x},y) = \left\{
\begin{array}{ll}
\Bra{\ve{x},y}\Oo{\Q}^\prime_\uparrow(E_0)\ket{\uparrow l};& y > 0,\\
\Bra{\ve{x},y}\Oo{\Q}^\prime_\downarrow(E_0)\ket{\downarrow l};& y < 0,
\end{array} \right. \label{eq:SOSes}
\end{equation}
$d_H(E_0)$, the maximal number of such linearly independent eigenfunctions
$\Psi_l(\ve{x},y)$ is at least $d_T(E_0) - d_K(E_0)$ since due to completeness
of SOS-eigenstates $\ket{n}$ the mapping $\Oo{Q}^\prime_\sigma(E_0)$ is
injective on the image $\op{K}^2(E_0){\cal L}$. But this number, $d_H(E_0)$,
can be larger than $d_T(E_0) - d_K(E_0)$ (but not larger than $d_T(E_0)$) since
there may be some states from the null space of $\op{K}^2(E_0)$ for which the
limits (\ref{eq:valQ},\ref{eq:derQ}) of the upper ($\sigma = \uparrow$) and
lower ($\sigma = \downarrow$) part accidentally match. Therefore we have
proved an inequality (\ref{eq:ineq}).

Analogously, one can show that the $d^\prime_T(E_0)$ basis vectors
$\ket{\downarrow n^*}\in {\cal L}, n = 1,\ldots,d^\prime_T(E_0)$ of the left
null space
\begin{equation}
\bra{\downarrow n^*}
\left(1 - \op{\T}_{\downarrow}(E_0)\op{\T}_\uparrow(E_0)\right) = 0.
\label{eq:qc3}
\end{equation}
are mapped onto conjugated basis of eigenfunctions under the propagator
$\oO{P}^\prime_{\sigma}(E_0)$
\begin{equation}
\Psi_n^*(\ve{x},y) = \left\{
\begin{array}{ll}
\bra{\uparrow n^*}\oO{P}^\prime_\uparrow(E_0)\Ket{\ve{x},y};
& y > 0,\\
\bra{\downarrow n^*}\oO{P}^\prime_\downarrow(E_0)\Ket{\ve{x},y};
& y < 0,
\end{array} \right. \label{eq:SOSes2}
\end{equation}
where the counterparts $\ket{\uparrow n^*}$ are again defined as
\begin{equation}
\bra{\uparrow n^*} = \bra{\downarrow n^*}\op{T}_\downarrow(E_0).
\end{equation}
Generally, these conjugated wavefunctions are continuous and
differentiable on the CSOS and are thus eigenfunctions of the Hamiltonian
$\Op{H}$ if $\ket{\sigma n^*}\in\op{K}^2(E_0){\cal L}$,
otherwise if $\ket{\sigma n^*}\in\ker\op{K}^2(E_0)$ the
continouity can be accidentally satisfied in the limit $E\rightarrow E_0$ if
both contributions (upper and lower) coincide. This happens when
the corresponding limits for $\ket{\sigma n}$ (\ref{eq:valQ},\ref{eq:derQ})
coincide since the two cases differ only by a complex conjugation.
So, the dimensions of left and right null space of
$1 - \op{T}_\downarrow(E_0)\op{T}_\uparrow(E_0)$ should be the same
$d_T(E_0) = d^\prime_T(E_0)$.
\\\\
The operator $\op{T}^\prime(E) = \op{T}_\downarrow(E)\op{T}_\uparrow(E)$
will be called {\em quantum Poincar\' e mapping} and it is the
product of the two {\em quantized Poincar\' e scattering mappings}.
We have proved an extremely efficient quantization condition (as we shall
show later), namely: The energies where the quantum Poincar\' e mapping
has fixed points (eigenvalue $1$) are either: (i) eigenenergies
of the Hamiltonian $\Op{H}$ or (ii) thresholds for opening of new
channels $E^\prime_n$ (which are already known as a solution of
(\ref{eq:eigenmodes}) as a prerequisite of the method).

\subsection{SOS decomposition of the resolvent of the Hamiltonian}

The kernels of the scattering propagators
$\bra{\ve{x}}\op{T}_\sigma(E)\ket{\ve{x}^\prime}$ will henceforth be called
CSOS-CSOS propagators. Then we define also:
(i) The linear operator $\Oo{Q}_\sigma(E)$ from ${\cal L}$ to
${\cal H}$ and the linear operator $\oO{P}_\sigma(E)$ from ${\cal H}$ to
${\cal L}$ with the kernels
\begin{eqnarray}
\Bra{\ve{x},y}\Oo{Q}_\sigma(E)\ket{\psi} &=&
\left\{ \begin{array}{ll}
\Bra{\ve{x},y}\Oo{Q}^\prime_\sigma(E)\ket{\psi}; & \sigma y \ge 0, \\
0; & \sigma y < 0, \end{array}\right. \label{eq:defQ} \\
\bra{\psi}\oO{P}_\sigma(E)\Ket{\ve{x},y} &=&
\left\{ \begin{array}{ll}
\bra{\psi}\oO{P}^\prime_\sigma(E)\Ket{\ve{x},y};
& \sigma y \ge 0, \\ 0; & \sigma y < 0. \end{array}\right. \label{eq:defP}
\end{eqnarray}
which are called CSOS-CS and CS-CSOS propagators respectively,
and (ii) a linear operator $\Op{G}_0(E)$ over ${\cal H}$ with the kernel
\begin{equation}
\Bra{\ve{x},y}\op{G}_0(E)\Ket{\ve{x}^\prime,y^\prime} =
\left\{ \begin{array}{ll}
\Bra{\ve{x},y}\op{G}_\uparrow(E)\Ket{\ve{x}^\prime,y^\prime};
& y \ge 0,\; y^\prime \ge 0,\\
\Bra{\ve{x},y}\op{G}_\downarrow(E)\Ket{\ve{x}^\prime,y^\prime};
& y \le 0,\; y^\prime \le 0,\\
0; & y y^\prime < 0. \end{array}
\right.
\label{eq:G0}
\end{equation}
which is called CS-CS propagator (without crossing the CSOS in between).
\\\\
{\bf Theorem 2a:}
{\em The energy-dependent quantum propagator (i.e. the resolvent of the
Hamiltonian) $\op{G}(E) = (E - \op{H})^{-1}$ can be decomposed in terms of the
CS--CS propagator --- with no intersection with the CSOS ${\cal S}_0$ ---
$\op{G}_0(E)$, CS--CSOS propagator $\oO{P}_\sigma(E)$, CSOS--CS
propagator $\Oo{\Q}_\sigma(E)$, and CSOS--CSOS propagator $\op{\T}_\sigma(E)$}
\begin{eqnarray}
\op{G}(E) = \op{G}_0(E) &+& \sum\limits_\sigma \Oo{Q}_\sigma(E)
(1-\op{\T}_{-\sigma}(E)\op{\T}_{\sigma}(E))^{-1} \oO{P}_{-\sigma}(E)
\nonumber \\
&+& \sum\limits_\sigma \Oo{Q}_{\sigma}(E)
(1-\op{\T}_{-\sigma}(E)\op{\T}_{\sigma}(E))^{-1}
\op{T}_{-\sigma}(E)\oO{P}_{\sigma}(E).
\label{eq:decomp}
\end{eqnarray}
\\\\
Quantities $\Bra{\ve{q}}\Op{G}_0(E)\Ket{\ve{q}^\prime},\,
\Bra{\ve{q}}\Oo{Q}_\sigma(E)\ket{\ve{x}^\prime},\,
\bra{\ve{x}}\oO{P}_\sigma(E)\Ket{\ve{q}^\prime},\,
\bra{\ve{x}}\op{T}_\sigma(E)\ket{\ve{x}^\prime}$
should be interpreted as the probability amplitudes to propagate through
the $\sigma$-side of CS from point $\ve{q}^\prime$ in CS / $\ve{x}^\prime$
on CSOS to point $\ve{q}$ in CS / $\ve{x}$ on CSOS at energy $E$ and without
crossing CSOS in between.
Then this decomposition formula can be intuitively understood by expanding the
operator $(1-\op{T}_{-\sigma}(E)\op{T}_\sigma(E))^{-1}$ in a geometric
series and then using the basic postulates of quantum mechanics
about summation of the probability amplitudes of alternative events
(different number of crossings of CSOS) and
multiplication of the probability amplitudes of consecutive events
(sequential crossings of CSOS) \cite{FH65}, since
the system which propagates from point $\ve{q}_i$ to point $\ve{q}_f$
in CS along continuous path can cross the CSOS arbitrarily many times.
(In fact, the number of crossings is {\em even} if
$\ve{q}_i$ and $\ve{q}_f$ lie on the same side of CSOS and {\em odd}
otherwise.)
Two versions of the proof of this formula are given in \cite{P94a,P94b}
while in this paper the proof will be given for more general cases
which include the present one in the following two sections.

\subsection{Semiclassical limit}

In order to find explicit leading order semiclassical expressions for
the CSOS/CS-CSOS/CS propagators it is convenient to express them first
in terms of the scattering Green functions in the hybrid representation
(\ref{eq:WFG},\ref{eq:G},\ref{eq:WFGc})
\begin{eqnarray*}
\op{T}_\sigma(E) &=& \frac{i\hbar^2}{m}\op{K}^{1/2}(E)
(\op{G}_\sigma(0,0,E)-\op{G}_{\rm free}(0,0,E))\op{K}^{1/2}(E),\\
\Bra{\ve{x},y}\Oo{Q}_\sigma(E) &=& \frac{\hbar}{\sqrt{-im}}\theta(\sigma y)
\bra{\ve{x}}\op{G}_\sigma(y,0,E)\op{K}^{1/2}(E),\\
\oO{P}_\sigma(E)\Ket{\ve{x},y} &=& \frac{\hbar}{\sqrt{-im}}\theta(\sigma y)
\op{K}^{1/2}(E)\op{G}_\sigma(0,y,E)\ket{\ve{x}},
\end{eqnarray*}
where the fourth propagator $\op{G}_0(y,y^\prime,E)$ is already
defined in terms of the scattering resolvents (\ref{eq:G0}) and
$\theta(y)$ is the well known Heaviside step function. Then define a linear
operator called {\em half-derivative} with the prescription
\begin{equation}
\partial^{1/2}_y e^{ay} = a^{1/2} e^{ay},\quad\re\,a^{1/2} \ge 0,
\label{eq:halfder}
\end{equation}
which is a sensibly defined positive square root of the
differential operator $\partial_y$. This is nonlocal operator which can be
explicitly expressed for functions $f(y)$ which increase slower that the
square root as $y$ goes towards plus or minus infinity,
$$
\partial^{1/2}_y f(y) = \frac{1}{\sqrt{-4\pi}}\int_y^\infty
d y^\prime \frac{f(y^\prime) - f(y)}{(y^\prime - y)^{3/2}}
\;\; {\rm or} \;\;
\partial^{1/2}_y f(y) = \frac{1}{\sqrt{4\pi}}\int_{-\infty}^y
d y^\prime \frac{f(y) - f(y^\prime)}{(y - y^\prime)^{3/2}},
$$
respectively. Thus for an operator valued exponential function,
$\partial^{1/2}_y e^{i\op{K} y} = \sqrt{i}\op{K}^{1/2} e^{i\op{K} y}$.
One may use the Schr\" odinger equation with proper boundary
conditions (which were used to derive eqs. (\ref{eq:WFG},\ref{eq:WFGc}))
to see that the scattering Green functions with one coordinate in the
waveguide may be written as exponential functions in that coordinate
\begin{equation}
\begin{array}{l}
\op{G}_\sigma(y,y^\prime,E)=e^{-i\sigma\op{K}(E) y}\op{f}(y^\prime,E),\\
\op{G}_\sigma(y^\prime,y,E)=\op{g}(y^\prime,E)e^{-i\sigma\op{K}(E) y},
\end{array}
\quad{\rm if}\;\; \sigma y \le 0,\;\sigma y^\prime > 0, \label{eq:Gin}
\end{equation}
for some operator valued functions $\op{f}$ and $\op{g}$. Using the
forms (\ref{eq:Gdiff},\ref{eq:Gin}) and the definition of half-derivative
one may rewrite the propagators in a more useful form
\begin{eqnarray}
\bra{\ve{x}}\op{T}_\sigma(E)\ket{\ve{x}^\prime} &=&
-\frac{\hbar^2}{\sigma m}\partial^{1/2}_y\partial^{1/2}_{y^\prime}
\Bra{\ve{x},y}\Op{G}_\sigma(E)-\Op{G}_{\rm free}(E)
\Ket{\ve{x}^\prime,y^\prime}\vert_{y=0}^{y^\prime=0},\label{eq:explT}\\
\Bra{\ve{x},y}\Oo{Q}_\sigma(E)\ket{\ve{x}^\prime} &=&
-\frac{\hbar\theta(\sigma y)}{\sqrt{\sigma m}}
\partial^{1/2}_{y^\prime}
\Bra{\ve{x},y}\Op{G}_\sigma(E)\Ket{\ve{x}^\prime,y^\prime}
\vert_{y^\prime = 0},\label{eq:explQ}\\
\bra{\ve{x}^\prime}\oO{P}_\sigma(E)\Ket{\ve{x},y} &=&
-\frac{\hbar\theta(\sigma y)}{\sqrt{\sigma m}}
\partial^{1/2}_{y^\prime}
\Bra{\ve{x}^\prime,y^\prime}\Op{G}_\sigma(E)\Ket{\ve{x},y}\vert_{y^\prime=0}.
\label{eq:explP}
\end{eqnarray}
{}From these formulae one can easily derive semiclassical approximations
by using the leading order semiclassical approximation for the energy
dependent Green function (see e.g. \cite{L90,B92})
\begin{eqnarray}
\Bra{\ve{q}}\Op{G}_\sigma(E)\Ket{\ve{q}^\prime} & \cong &
\frac{2\pi}{(2\pi i \hbar)^{(f+1)/2}}\sum\limits_j B_j(\ve{q},\ve{q}^\prime,E)
e^{i S_j(\sve{q},\sve{q}^\prime,E)/\hbar - i\nu_j\pi/2},\label{eq:scG}\\
B_j(\ve{q},\ve{q}^\prime,E) &=&
\left\arrowvert\det\pmatrix{
\partial_{\ve{q}}\partial_{\ve{q}^\prime} S_j &
\partial_{\ve{q}}\partial_E S_j \cr
\partial_E\partial_{\ve{q}^\prime} S_j &
\partial^2_E S_j \cr}\right\arrowvert^{1/2} =
m \frac{D_j(\ve{q},\ve{q}^\prime,E)}
{\left\arrowvert p_{yj}p^\prime_{yj}\right\arrowvert^{1/2}},\nonumber\\
D_j(\ve{q},\ve{q}^\prime,E) &=&
\left\arrowvert\det\partial_{\ve{x}}\partial_{\ve{x}^\prime}
S_j(\ve{q},\ve{q}^\prime,E)\right\arrowvert^{1/2},\nonumber
\end{eqnarray}
where the sum is taken over ({\em usually finitely many}) classical
scattering trajectories labeled by $j$ with classical actions
$S_j(\ve{q},\ve{q}^\prime,E) = \int_j d\ve{q}\cdot\ve{p}$, and Morse
indices $\nu_j$ which count the number of conjugated points along the orbit
$j$.
$p_{yj} = \partial_y S_j$ and $p^\prime_{yj} = -\partial_{y^\prime} S_j$
are the perpendicular (w.r.t. CSOS) projections of the final and
initial momenta. Thus using the definition (\ref{eq:halfder}) in the
{\em leading semiclassical order} the half derivatives only cancel the square
roots of $y$-momenta if one expresses root of $(f+1)\times (f+1)$ determinant
$B_j$ in terms of root of $(f-1)\times (f-1)$ determinant $D_j$,
\begin{eqnarray}
\bra{\ve{x}}\op{T}_\sigma(E)\ket{\ve{x}^\prime} & \cong &
\frac{1}{(2\pi i \hbar)^{(f-1)/2}}\sum\limits_j^\prime
D_j(\ve{x},0,\ve{x}^\prime,0,E)
e^{i S_j(\sve{x},0,\sve{x}^\prime,0,E)/\hbar - i\nu_j\pi/2},\quad\quad
\label{eq:scT}\\
\Bra{\ve{x},y}\Oo{Q}_\sigma(E)\ket{\ve{x}^\prime} & \cong &
\frac{i\sqrt{2\pi m}\theta(\sigma y)}{(2\pi i \hbar)^{f/2}}\sum\limits_j
\frac{D_j(\ve{x},y,\ve{x}^\prime,0,E)}
{\left\arrowvert p^\prime_{yj}\right\arrowvert^{1/2}}
e^{i S_j(\sve{x},y,\sve{x}^\prime,0,E)/\hbar - i\nu_j\pi/2},\quad\quad
\label{eq:scQ}\\
\bra{\ve{x}^\prime}\oO{P}_\sigma(E)\Ket{\ve{x},y} & \cong &
\frac{i\sqrt{2\pi m}\theta(\sigma y)}{(2\pi i \hbar)^{f/2}}\sum\limits_j
\frac{D_j(\ve{x}^\prime,0,\ve{x},y,E)}
{\left\arrowvert p^\prime_{yj}\right\arrowvert^{1/2}}
e^{i S_j(\sve{x}^\prime,0,\sve{x},y,E)/\hbar - i\nu_j\pi/2}.\quad\quad
\label{eq:scP}
\end{eqnarray}
In (\ref{eq:scT}) the sum $\sum^\prime_j$ is restricted only to classical
orbits which strictly leave CSOS and lie entirely on the $\sigma$-side of CS.
The trivial classical scattering orbit whose
$y$-coordinate is constantly zero is the only classical orbit of the
semiclassical {\em free} Green function $\Op{G}_{\rm free}(E)$ and is thus
canceled in the expression (\ref{eq:explT}) for the propagator
$\op{T}_\sigma(E)$. The sums in (\ref{eq:scQ},\ref{eq:scP}) contain
only the classical orbits which lie entirely on the $\sigma-$side of
the CS with one end point on the CSOS and the other end point lying in
the $\sigma-$side of CS. If there was a classical orbit whose part
would lie in the waveguide then its $y$-coordinate should have an
extremum there which, however, is impossible since the classical motion in
the waveguide is free in $y$-direction. The semiclassical expression
for the CS-CS propagator without crossing the CSOS
$\Bra{\ve{x},y}\Op{G}_0(E)\Ket{\ve{x}^\prime,y^\prime}$
thus according to definition (\ref{eq:G0}) looks the same as
RHS of (\ref{eq:scG}) where the sum now includes only the classical orbits
which {\em do not cross} CSOS. The leading order asymptotic results
(\ref{eq:scT},\ref{eq:scQ},\ref{eq:scP})
agree with the semiclassical theory of Bogomolny \cite{B92}.
\\\\
The {\em higher order semiclassical corrections} to CSOS/CS-CSOS/CS
propagators can be obtained in a systematic way
by (i) inserting corrected higher order semiclassical expression for the
scattering Green function $(\ref{eq:scG})$ in the formulae (\ref{eq:explT},
\ref{eq:explQ}, \ref{eq:explP}) and (ii) evaluating the half-derivatives in
terms of a power series in $\hbar$. I will show now briefly how both steps
can be systematically performed.
\\\\
(i) {\em The higher order corrections to semiclassical energy-dependent
Green function} $\Bra{\ve{q}}\Op{G}_\sigma(E)\Ket{\ve{q}^\prime}$ can be
obtained \footnote{A variant of this approach for time-dependent
quantum propagator (without consideration of conjugated points
-- short time limit) has been developed by Roncadelli \cite{R94}, whereas
Gaspard and Alonso \cite{GA93} used another (path integral) approach to derive
$\hbar$-expansion of Gutzwiller trace formula.}
by multiplying each term of the formula (\ref{eq:scG}) by a correction factor
$$
\Bra{\ve{q}}\Op{G}_\sigma(E)\Ket{\ve{q}^\prime} =
\frac{2\pi}{(2\pi i \hbar)^{(f+1)/2}}\sum\limits_j
B_j e^{i S_j/\hbar - i\nu_j\pi/2}\sum_{n=0}^\infty\hbar^n
f^j_n(\ve{q},\ve{q}^\prime,E).
$$
The corrections $f^j_n$ can be calculated by inserting the whole
expression to the Schr\" odinger equation. Comparing the
terms with equal power of $\hbar$ and integrating along the orbit
yields the explicit recursion formulae for the semiclassical corrections
\begin{eqnarray*}
f^j_0(\ve{q}_j(t),\ve{q}^\prime,E) &=& 1, \nonumber\\
f^j_n(\ve{q}_j(t),\ve{q}^\prime,E) &=& \frac{i}{2m}
(-)^{n\nu_j(\sve{q}_j(t),\sve{q}^\prime,E)}\;{\cal P}\!\!
\int_0^t dt^\prime
(-)^{n\nu_j(\sve{q}_j(t^\prime),\sve{q}^\prime,E)}
\Op{\Delta}_j f^j_{n-1}(\ve{q}_j(t^\prime),\ve{q}^\prime,E),\\
\Op{\Delta}_j f(\ve{q},\ve{q}^\prime,E) &=&
B^{-1}_j(\ve{q},\ve{q}^\prime,E)
\partial^2_{\ve{q}}\left[
B_j(\ve{q},\ve{q}^\prime,E) f(\ve{q},\ve{q}^\prime,E)\right],
\end{eqnarray*}
where $\ve{q}_j(t)$ denotes the classical orbit $j$ with end
points $\ve{q}^\prime$ and $\ve{q}$.
One must use the sign factors
$(-)^{n\nu_j}$ and the Cauchy principal value of the
integral in order to avoid infinite contributions
each time one passes a singularity -- conjugated point.
\\\\
(ii) Half derivative of a term like $e^{iS/\hbar} f$, where we shall take
$S = S_j,\, f = B_j f^j_n$, may be represented as a power series
$$
\partial^{1/2}_y e^{iS/\hbar} f = \frac{1}{\sqrt{\hbar}}e^{iS/\hbar}
\sum_{n=0}^\infty\hbar^n\Op{C}_n f
$$
where $\Op{C}_n$ are some linear operators independent of $\hbar$.
Taking it twice, $\partial^{1/2}_y\partial^{1/2}_y = \partial_y$,
and comparing the terms with the same power of $\hbar$ one obtains
the set of equations which determine the operators $\Op{C}_n$
\begin{eqnarray*}
\Op{C}^2_0 &=& i \partial_y S, \\
\Op{C}_0\Op{C}_1 + \Op{C}_1\Op{C}_0 &=& \partial_y, \\
\sum\limits_{m=0}^n \Op{C}_m\Op{C}_{n-m} &=& 0,\quad n\ge 2.
\end{eqnarray*}
It is easy to see that $\Op{C}_n$ is $n-$th order differential operator.
For example, we give explicit expressions for the first two
\begin{equation}
\Op{C}_0 = (i\partial_y S)^{1/2},\quad
\Op{C}_1 = (i\partial_y S)^{1/2}
\left(\frac{i\partial^2_y S_j}{8(\partial_y S)^2} -
      \frac{i}{2\partial_y S}\partial_y\right)
\end{equation}
and the {\em next-to-leading order semiclassical} expression for the quantum
CSOS-CSOS propagator
\begin{eqnarray}
\bra{\ve{x}}\op{T}_\sigma(E)\ket{\ve{x}^\prime} & \cong &
\frac{1}{(2\pi i \hbar)^{(f-1)/2}}\sum\limits_j^\prime
D_j e^{i S_j/\hbar - i\nu_j\pi/2} \\
&\times& \left[1 + \hbar\left(f^j_1 +
\frac{3 i\partial^2_y S_j}{8 p^2_{yj}} +
\frac{3 i\partial^2_{y^\prime} S_j}{8 p^{\prime 2}_{yj}} -
\frac{i\partial_y D_j}{2 p_{yj}D_j} +
\frac{i\partial_y D_j}{2 p^\prime_{yj}D_j}\right)\right] \nonumber
\end{eqnarray}
where all functions on RHS have arguments $(\ve{x},0,\ve{x}^\prime,0,E)$.

\subsection{Symmetry of the CSOS-CSOS propagator}

At a given value of energy $E$ one can split the SOS-Hilbert space on two
orthogonal components,
$$ {\cal L} = {\cal L}_{\rm o}(E) \oplus {\cal L}_{\rm c}(E), $$
the (usually finite dimensional) subspace of open channels
and the (usually infinitely dimensional) subspace of closed channels
\begin{eqnarray*}
{\cal L}_{\rm o}(E) &=& \big\{\ket{\psi}\in {\cal L},\,
\bra{\psi}\op{K}^2(E)\ket{\psi} \ge 0\big\},\\
{\cal L}_{\rm c}(E) &=& \big\{\ket{\psi}\in {\cal L},\,
\bra{\psi}\op{K}^2(E)\ket{\psi} < 0\big\}
\end{eqnarray*}
spanned by $\{\ket{n},E_n^\prime \le E\}$, and by $\{\ket{n},E_n^\prime > E\}$,
respectively.

A very useful information about the CSOS-CSOS propagator which can be written
in a block form
$$\op{T}_\sigma(E) = \pmatrix{
\op{T}_{\rm oo}(E) & \op{T}_{\rm oc}(E) \cr
\op{T}_{\rm co}(E) & \op{T}_{\rm cc}(E) \cr}$$
can be obtained by comparing the two expressions for
scattering wavefunctions, conjugated eq. (\ref{eq:scWFc}) and eq.
(\ref{eq:scWF}). Comparing the values and normal derivatives on CSOS one
obtains two equations
$$
\sqrt{\mp i}\op{K}^{\mp 1/2}(1 \pm \op{T}_\sigma)\ket{\psi} =
\sqrt{\pm i}\op{K}^{\mp 1/2 \dagger}(1 \pm \op{T}^\dagger_\sigma)\ket{\psi^*}.
$$
Noting that
$\op{K}^{\pm 1/2}\op{K}^{\mp 1/2 \dagger} = \pmatrix{1 & 0 \cr 0 & \pm i\cr}$
and performing some algebra yields
\begin{eqnarray}
\op{T}_{\rm oo}\op{T}^\dagger_{\rm oo} &=&
\op{T}^\dagger_{\rm oo}\op{T}_{\rm oo} = 1,\label{eq:uni}\\
i\op{T}_{\rm oo}\op{T}^\dagger_{\rm co} &=& \op{T}_{\rm oc},\\
i\op{T}^\dagger_{\rm oc}\op{T}_{\rm oo} &=& \op{T}_{\rm co},\\
i\op{T}_{\rm co}\op{T}^\dagger_{\rm co} &=&
i\op{T}^\dagger_{\rm oc}\op{T}_{\rm oc} =
\op{T}_{\rm cc} - \op{T}^\dagger_{\rm cc}.
\end{eqnarray}
This is the so called {\em generalized unitarity} \cite{SS93,RS94} of a
CSOS-CSOS propagator. Note that the open-open part $\op{T}^{\rm oo}_\sigma(E)$
is indeed a unitary operator (\ref{eq:uni}).

Now we give the representation of the CSOS-CSOS propagator in terms
of what is called {\em reactance matrix} in scattering theory \cite{N82}.
This gives also a practical recipe to determine the CSOS-CSOS
propagators $\op{T}_\sigma(E)$. Let $\Psi_{\sigma n}(\ve{x},y,E)$ denote
the unique wavefunctions which satisfy Schr\" odinger equation
(\ref{eq:Schreq}) with boundary conditions
$$\Psi_{\sigma n}(\ve{x},0,E) = \braket{\ve{x}}{n},\quad
  \Psi_{\sigma n}(\ve{x},\sigma\infty,E) = 0, $$
where the second condition should be taken on the boundary of CS if
the latter is not infinite. Then one can define the {\em reactance operators}
$\op{R}_\sigma(E)$ with matrix elements
\begin{equation}
\bra{l}\op{R}_\sigma(E)\ket{n} = \frac{\sigma}{k^{1/2}_l(E) k^{1/2}_n(E)}
\int\limits_{\cal S} d\ve{x} \Psi^*_{\sigma l}(\ve{x},y,E)\partial_y
\Psi_{\sigma n}(\ve{x},y,E)\vert_{y=0}
\label{eq:defR}
\end{equation}
and show (using (\ref{eq:scWF})) that the CSOS-CSOS propagators can be
written as
\begin{equation}
\op{T}_\sigma(E) = (1 + i\op{R}_\sigma(E))(1 - i\op{R}_\sigma(E))^{-1}.
\label{eq:Treac}
\end{equation}
In case of time reversal symmetry the wave-functions
$\Psi_{\sigma n}(\ve{x},y,E)$ are real and one can use Green's theorem to
show that then the reactance matrix is symmetric
$$
\bra{l}\op{R}_\sigma(E)\ket{n} = \bra{n}\op{R}_\sigma(E)\ket{l}
\quad {\rm if}\;\; \Psi_{\sigma n}(\ve{x},y,E) = \Psi^*_{\sigma n}(\ve{x},y,E).
$$
Using representation (\ref{eq:Treac}) this means that then also the CSOS-CSOS
propagator is symmetric
\begin{equation}
\bra{l}\op{T}_\sigma(E)\ket{n} = \bra{n}\op{T}_\sigma(E)\ket{l},\quad
{\rm or}\quad
\bra{\ve{x}}\op{T}_\sigma(E)\ket{\ve{x}^\prime} =
\bra{\ve{x}^\prime}\op{T}_\sigma(E)\ket{\ve{x}}.
\end{equation}

\subsection{Practical applications and semi-separable systems}

Let us truncate the basis of ${\cal L}$ to include all $N_o$ open channels
of ${\cal L}_o$ and the first $N_c$ closed channels.
The truncated $(N = N_o + N_c)-$dim matrices with matrix elements
$\bra{l}\op{T}_\sigma(E)\ket{n}$ and
$\bra{l}\op{R}_\sigma(E)\ket{n}$ will be denoted by
$\ma{T}_\sigma(E)$ and $\ma{R}_\sigma(E)$, respectively.
In practice one should increase $N_c$ until numerical results converge, which
is typically the case \cite{SS93,RS94,P94c} for very small values of $N_c$
already, so in the semiclassical limit $N\approx N_o$. The practical
SOS-quantization condition then reads
\begin{equation}
\det(1-\ma{T}_\downarrow(E)\ma{T}_\uparrow(E)) = 0.
\label{eq:qcT}
\end{equation}
Using (\ref{eq:Treac}) this condition can be formulated in terms of
reactance matrices
\begin{equation}
\det(\ma{R}_\uparrow(E) + \ma{R}_\downarrow(E)) = 0.
\label{eq:qcR1}
\end{equation}
In case of systems having a time-reversal symmetry
(that is if $H^\prime(\ve{p}_{\ve{x}},\ve{x},y) =
H^\prime(-\ve{p}_{\ve{x}},\ve{x},y)$) the reactance matrix (\ref{eq:defR}) is a
complex symmetric matrix $\ma{R}_\sigma(E) = \ma{R}^T_\sigma(E)$ which is
written in terms of a purely real matrix $\tilde{\ma{R}}_\sigma(E)$ via
$$
\ma{R}_\sigma(E) = \pmatrix{ 1 & 0 \cr 0 &\sqrt{-i}\cr}
\tilde{\ma{R}}_\sigma(E)\pmatrix{ 1 & 0 \cr 0 &\sqrt{-i}\cr}
$$
where the diagonal elements are $N_o$ and $N_c$ dimensional sub-matrices.
The quantization condition (\ref{eq:qcR1}) can be thus expressed in terms of
{\em purely real symmetric} matrices $\tilde{\ma{R}}_\sigma(E) =
\tilde{\ma{R}}^T_\sigma(E) = \tilde{\ma{R}}^*_\sigma(E)$
\begin{equation}
\det(\tilde{\ma{R}}_\uparrow(E) + \tilde{\ma{R}}_\downarrow(E)) = 0.
\label{eq:qcR}
\end{equation}
The equation (\ref{eq:qcR}) is much more efficient for numerical calculation
of energy spectra (by seeking its zeros) than original quantization
condition (\ref{eq:qcT}) since the former involves real arithmetic
\cite{P94c}.

There is a generic (in a sense of dynamics) class of the so called
{\em semi-separable} systems for which reactance matrices can be
straightforwardly calculated and hence quantization condition (\ref{eq:qcR})
can be easily implemented. Semi-separable system should be separable
(in $(\ve{x},y)$ coordinates) on both sides of CSOS (for $y > 0$ and $y < 0$)
but it can be discontinuous on CSOS ($\op{H}^\prime(-0)\ne\op{H}^\prime(+0)$).
Thus one has two complete sets of {\em normalized} SOS eigenmodes, first
$\ket{n}_+$ are eigenstates of $\op{H}^\prime(+0)$, and the second $\ket{n}_-$
are eigenstates of $\op{H}^\prime(-0)$.
Since the system is separable on both sides one can
explicitly calculate the wavefunctions $\Psi_{\sigma n}(\ve{q},E)$ by
separation of coordinates
\begin{eqnarray*}
\Psi_{\uparrow n}(\ve{x},y,E) &=&
\phi_{\uparrow n}(y,E)\braket{\ve{x}}{n}_+, \quad y > 0,\\
\Psi_{\downarrow n}(\ve{x},y,E) &=&
\phi_{\downarrow n}(y,E)\braket{\ve{x}}{n}_-, \quad y < 0,
\end{eqnarray*}
where $y$-dependent parts should be normalized to give $\phi_{\sigma n}(0) =
1$.
We have freedom to cut CS slightly above discontinuity and
choose priveliged set $\ket{n}_+$ with wavenumbers $k_n(E)$.
Then we apply the definition (\ref{eq:defR}) to calculate real reactance
matrices
\begin{eqnarray*}
\ma{R}_{\uparrow nl}(E) &=& \vert k_n(E)\vert^{-1}
\partial_y\phi_{\uparrow n}(0,E)\delta_{nl},\\
\ma{R}_{\downarrow nl}(E) &=& -\vert k_n(E) k_l(E)\vert^{-1/2} \sum_{j}
\ _+\braket{n}{j}_- \,\partial_y\phi_{\downarrow j}(0,E)\ _-\braket{j}{l}_+.
\end{eqnarray*}
The upper is diagonal while the lower includes transformations
by means of real orthogonal matrix with elements
$_-\braket{l}{n}_+ =\ _+\braket{n}{l}_-$
which are typically easily calculated knowing the two sets of SOS-eigenmodes.
The author has applied this method for numerical calculation of energy
levels in the so called {\em semi-separable 2-dim oscillator} \cite{P94c}
which is a {\em generic nonlinear autonomous dynamical system with two
freedoms}. The method turned out to be capable of yielding consecutive energy
levels with sequential number of the order of $10^6$ (million !) at the cost
of few minutes of Convex C3860 CPU time per level.

\section{Abstract formulation of the method}

In this section we devise a {\em general} and {\em abstract}
mathematical framework within which one can prove all versions of simply
connected SOS quantization conditions and SOS decompositions of the
resolvents of the corresponding Hamilton operators.

Let ${\cal M}$ be an arbitrary normed vector space, which will be referred to
as
{\em reduced space}. The vectors from reduced space ${\cal M}$ --
{\em r-vectors} will be written in bold and linear operators over
reduced space -- {\em r-operators} will wear mathematical accent $\op{}$.
Then we define r-operator valued function $\op{\ma{L}}(y,E)$, where $y$ is a
real variable and $E$ is a complex {\em spectral parameter} (e.g. energy), in
order to study the following general homogeneous vector differential equation
\begin{equation}
\left(\partial_y - \op{\ma{L}}(y,E)\right)\ve{f}(y) = 0
\label{eq:homeq}
\end{equation}
on the entire real axis $y\in {\cal R}$ (or on some finite interval
which contains zero). Normalized r-vector valued functions $\ve{f}(y)$,
$\int dy \Vert\ve{f}(y)\Vert^2 < \infty$,
constitute a normed vector space, denoted by ${\cal H}$. The values of spectral
parameter $E$ for which the equation (\ref{eq:homeq}) has nontrivial
solutions in ${\cal H}$ are called {\em generalized eigenvalues} whereas the
corresponding r-vector valued functions are called {\em generalized
eigenfunctions} of the equation (\ref{eq:homeq}). It will be shown in the next
subsection that this problem is equivalent to time-independent
Schr\" odinger equation for some special choice of $\op{\ma{L}}(y,E)$.
The equation (\ref{eq:homeq}) can be generally solved by means of a
Green function $\op{\ma{G}}(y,y^\prime,E)$ which is a unique r-operator valued
function which solves the inhomogeneous equation
\begin{equation}
\left(\partial_y - \op{\ma{L}}(y,E)\right)\op{\ma{G}}(y,y^\prime) =
\delta(y-y^\prime)\op{\ma{J}}
\label{eq:inheq}
\end{equation}
with boundary conditions
\begin{equation}
\lim\limits_{y\rightarrow\pm\infty}\op{\ma{G}}(y,y^\prime,E) = 0.
\label{eq:bc}
\end{equation}
where $\op{\ma{J}}$ can be an arbitrary nonsingular r-operator.

Equation (\ref{eq:inheq}) may be written in the form $\Op{L}(E)\Op{G}(E) = 1$
where $\Op{L}(E)$ and $\Op{G}(E)$ are the operators over ${\cal H}$ with
kernels being r-operator valued distributions
$\op{\ma{J}}^{-1}(\delta^\prime(y-y^\prime) -
\op{\ma{L}}(y,E)\delta(y - y^\prime))$ and
$\op{\ma{G}}(y,y^\prime,E)$, respectively. If left and right inverse
of $\Op{L}(E)$ exist and coincide then $\Op{G}(E)\Op{L}(E) = 1$, so the
Green function satisfies also the ``conjugated'' equation
\begin{equation}
\op{\ma{G}}(y,y^\prime)\left(\overleftarrow{\partial}_{y^\prime} -
\op{\ma{L}}^\prime (y^\prime,E)\right) = -\delta(y - y^\prime)\op{\ma{J}}.
\end{equation}
where
$$\op{\ma{L}}^\prime(y,E) = -\op{\ma{J}}^{-1}\op{\ma{L}}(y,E)\op{\ma{J}}.$$

We shall construct the Green function $\op{\ma{G}}(y,y^\prime,E)$ by means of
the Green functions $\op{\ma{G}}_{\sigma}(y,y^\prime,E)$ of two {\em
generalized
scattering} problems which are defined by cutting the $y-$axis at $y=0$ and
substituting the upper ($y > 0, \sigma=\uparrow = + $)/lower ($y < 0,
\sigma=\downarrow = - $) part of function $\op{\ma{L}}(y,E)$ by a constant
$\op{\ma{L}}(E) = \op{\ma{L}}(0,E)$. Therefore these scattering Green functions
satisfy
\begin{eqnarray}
\left(\partial_y - \op{\ma{L}}(y,E)\right)\op{\ma{G}}_{\sigma}(y,y^\prime)
&=& \delta(y-y^\prime)\op{\ma{J}},\quad{\rm if}\;\;\sigma y > 0,\\
\left(\partial_y - \op{\ma{L}}(0,E)\right)\op{\ma{G}}_{\sigma}(y,y^\prime)
&=& \delta(y-y^\prime)\op{\ma{J}},\quad{\rm if}\;\;\sigma y \le 0,
\label{eq:inhsceq}
\end{eqnarray}
with boundary conditions
\begin{equation}
\lim\limits_{y\rightarrow\sigma\infty}\op{\ma{G}}_{\sigma}(y,y^\prime,E) = 0.
\end{equation}
The scattering Green function $\op{\ma{G}}_{\sigma}(y,y^\prime,E)$ can be
written explicitly on $(-\sigma)-$side ($\sigma y \le 0, \sigma y^\prime \le
0$) in terms of the {\em abstract scattering operator}
$\op{\ma{T}}_{\sigma}(E)$
\begin{equation}
\op{\ma{G}}_{\sigma}(y,y^\prime,E) =
i\exp(\op{\ma{L}}(E) y)\left(
\half - \ihalf\sgn{y-y^\prime}\op{\ma{J}}
- \op{\ma{T}}_{\sigma}(E)\right)\exp(\op{\ma{L}}^\prime(E) y^\prime),
\label{eq:scatG}
\end{equation}
where $\sgn{y} = \uparrow = +; y > 0, \sgn{y} = \downarrow = -; y < 0$
denotes the {\em side} or {\em sign}. The equation (\ref{eq:scatG})
can be considered also as a unique definition of the abstract scattering
operator, or
\begin{equation}
\op{\ma{T}}_\sigma(E) = i\op{\ma{G}}_\sigma(\pm 0,\mp 0,E) + \half \mp
\ihalf\op{\ma{J}}.
\end{equation}
\\\\
{\bf Theorem 2b:}
{\em The Green function of the equation (\ref{eq:inheq})
$\op{\ma{G}}(y,y^\prime,E)$ can be
decomposed in terms of four r-operator valued functions which can be defined by
means of the scattering Green functions}
\begin{eqnarray}
\op{\ma{G}}_0(y,y^\prime,E) &=&
\delta_{\sgn{y}\sgn{y^\prime}} \op{\ma{G}}_{\sgn{y}}(y,y^\prime,E),\\
\op{\ma{Q}}(y,E) &=& \sqrt{i}\sgn{y} \op{\ma{G}}_{\sgn{y}}(y,0,E),\\
\op{\ma{P}}(y,E) &=& \sqrt{i}\sgn{y} \op{\ma{G}}_{\sgn{y}}(0,y,E),\\
\op{\ma{T}}(E) &=& \op{\ma{T}}_\uparrow(E) + \op{\ma{T}}_\downarrow(E)
= i\op{\ma{G}}_\uparrow(\pm 0,\mp 0,E) +
  i\op{\ma{G}}_\downarrow(\mp 0,\pm 0,E) + 1.
\end{eqnarray}
{\em Namely, the decomposition formula reads}
\begin{equation}
\op{\ma{G}}(y,y^\prime,E) = \op{\ma{G}}_0(y,y^\prime,E) +
\op{\ma{Q}}(y,E)\left(1 - \op{\ma{T}}(E)\right)^{-1}\op{\ma{P}}(y^\prime,E).
\label{eq:adecomp}
\end{equation}
\\\\
{\bf Proof:} One must show that RHS of (\ref{eq:adecomp}) is the
solution of equations (\ref{eq:inheq},\ref{eq:bc}). The first term
of RHS is indeed a solution of a inhomogeneous equation (\ref{eq:inheq})
with boundary conditions (\ref{eq:bc}) on both sides but it is
generally discontinuous at $y=0$. The second term of RHS, or its
first factor $\op{\ma{Q}}(y,E)$, is a solution of the homogeneous equation
(\ref{eq:homeq}) on both sides but it is again discontinuous at $y=0$.
The sum of the two terms is therefore also the solution of the
inhomogeneous equation $(\ref{eq:inheq})$. One is left to prove that the
sum is continuous at $y=0$ and therefore a unique solution of
equation (\ref{eq:inheq}).
For arbitrary function of $y$ we define the difference operator
\begin{equation}
\Delta_y f(y) = f(y + 0) - f(y - 0).
\label{eq:diffop}
\end{equation}
Then the straightforward calculation yields
\begin{eqnarray*}
\Delta_y \op{\ma{G}}(0,y^\prime,E) &=&
\Delta_y \op{\ma{G}}_0(0,y^\prime,E) + (\Delta_y \op{\ma{Q}}(0,E))
(1 - \op{\ma{T}}(E))^{-1}\op{\ma{P}}(y^\prime,E)= \\
&=& \Delta_y \op{\ma{G}}_0(0,y^\prime,E) -
\sgn{y^\prime}\op{\ma{G}}_{\sgn{y^\prime}}(0,y^\prime,E) = 0,
\end{eqnarray*}
since
\begin{equation}
\Delta_y \op{\ma{Q}}(0,E) = \sqrt{i}(1 - \op{\ma{T}}(E)).
\label{eq:diffq}
\end{equation}
\\\\
{\bf Theorem 1b:} {\em For any generalized eigenvalue $E$ of the equation
(\ref{eq:homeq}) the operator $\op{\ma{T}}(E)$ has a fixed point.}
\\\\
{\bf Proof:}
Any generalized eigenfunction of the equation $(\ref{eq:homeq})$
corresponding to generalized eigenvalue $E$ can be written in a form
\begin{equation}
\ve{f}(y) = \op{\ma{Q}}(y,E)\ve{a}
\label{eq:fy}
\end{equation}
for some nonzero r-vector $\ve{a}\in {\cal M}$. One can write explicitly,
$\ve{a} = \op{\ma{Q}}^{-1}(0,E)\ve{f}(0)$ if $\op{\ma{Q}}(0,E)$ is invertible,
or more generally,
$\ve{a} = [\partial^r_y\op{\ma{Q}}(y,E)]^{-1}\partial^r_y\ve{f}(y)|_{y=0}$
if $\partial^p_y\op{\ma{Q}}(y,E)|_{y=0}$ is singular for all
$p = 0,1\ldots r-1$. Eq. (\ref{eq:fy}) follows from the
definition of r-operator valued function $\op{\ma{Q}}(y,E)$ in terms of
scattering Green functions on the {\em nontrivial side}. Since the function
$\ve{f}(y)$ is continuous at $y=0,\; \Delta_y\ve{f}(0) = 0$, one
sees, using equation (\ref{eq:diffq}), that $\ve{a}$ is a fixed point of the
operator $\op{\ma{T}}(E)$,
$$\op{\ma{T}}(E)\ve{a} = \ve{a}.$$

Note that general decomposition formula (\ref{eq:adecomp}) is invariant
with respect to similarity trasnformations
\begin{equation}
\begin{array}{rcl}
\op{\ma{Q}}(y,E) & \rightarrow & \op{\ma{Q}}(y,E)\op{\ma{S}},\\
\op{\ma{T}}(E) & \rightarrow & \op{\ma{S}}^{-1}\op{\ma{T}}(E)\op{\ma{S}},\\
\op{\ma{P}}(y,E) & \rightarrow & \op{\ma{S}}^{-1}\op{\ma{P}}(y,E)
\end{array}
\label{eq:sim}
\end{equation}
and transformations
\begin{equation}
\begin{array}{rcl}
\op{\ma{Q}}(y,E) & \rightarrow & \op{\ma{Q}}(y,E)\op{\ma{Z}}^{-1},\\
\op{\ma{T}}(E) - 1 & \rightarrow & \op{\ma{Z}}(\op{\ma{T}}(E) -
1)\op{\ma{Z}},\\
\op{\ma{P}}(y,E) & \rightarrow & \op{\ma{Z}}^{-1}\op{\ma{P}}(y,E)
\end{array}
\label{eq:sim2}
\end{equation}
where $\op{\ma{S}},\op{\ma{Z}}$ are any bijective r-operators.

\subsection{Trivial application}

For example, let us first cast our ordinary Schr\" odinger
problem (\ref{eq:Schreq},\ref{eq:qH}) into the abstract form.
Reduced space should be here ${\cal M} = {\cal L}\oplus{\cal L}$,
since the Schr\" odinger equation is of the second order.
One should take
\begin{equation}
\op{\ma{L}}(y,E) = \pmatrix{ 0 & -1\cr \op{K}^2(y,E) & 0\cr},\quad
\op{\ma{J}} = \pmatrix{ 0 & -1\cr 1 & 0\cr}
\label{eq:applic}
\end{equation}
in (\ref{eq:homeq},\ref{eq:inheq}) where
$\op{K}^2(y,E) = (2m/\hbar^2)(E - \op{H}^\prime(y)),\,
\op{K}(E) = \op{K}(0,E)$. Then, referring to the two components ${\cal L}$
of ${\cal M}$ with indices $1$ and $2$, $\ket{\psi(y)} = \ve{f}_1(y)$
is a solution of the Schr\" odinger equation
$(\partial^2_y + \op{K}^2(y,E))\ket{\psi(y)} = 0$ and
$\op{G}(y,y^\prime,E)=-(2m/\hbar^2)\op{\ma{G}}_{11}(y,y^\prime,E)\op{K}^{-1}(E)$
is its normal Green function (\ref{eq:hybG}) in hybrid representation.
Comparing scattering Ans\" atze (\ref{eq:G}) and (\ref{eq:scatG}) and using
the similarity transformation (\ref{eq:sim})
$$\op{\ma{S}} = \pmatrix{\op{K}^{-1/2}(E) & \op{K}^{-1/2}(E)\cr
                         i\op{K}^{1/2}(E) &-i\op{K}^{1/2}(E)\cr}$$
one obtains
\begin{equation}
\op{\ma{S}}^{-1}\op{\ma{T}}(E)\op{\ma{S}} =
\pmatrix{ 0 & \op{T}_\uparrow(E) \cr \op{T}_\downarrow(E) & 0\cr}.
\label{eq:TT}
\end{equation}
One can also check that the first ``row'' of
$(i\hbar/\sqrt{2m})\op{\ma{Q}}(y,E)\op{\ma{S}}$ and the first ``column'' of
$(i\hbar/\sqrt{2m})\op{\ma{S}}^{-1}\op{\ma{P}}(y,E)\op{K}(E)$ can be written
as $(\Oo{Q}_\downarrow(E),\Oo{Q}_\uparrow(E))$ and
   $(\oO{P}_\uparrow(E),\oO{P}_\downarrow(E))^T$, respectively.
It is now easy to check that $1,1$ component of the general decomposition
formula (\ref{eq:adecomp}) agrees with the more special case (\ref{eq:decomp}).

\subsection{Nontrivial applications}

There is a straightforward nontrivial generalization of application
(\ref{eq:applic}), namely, one can include nonrelativistic systems which
interact with very general external (gauge) fields and have thus Hamiltonians
in our canonical coordinates $(\ve{x},y)$ of the form
\begin{equation}
\Op{H} = \frac{1}{2m}(-i\hbar\partial_y - \Op{A}(y))^2 + \Op{H}^\prime(y),
\label{eq:hamA}
\end{equation}
where the only restriction for the self-adjoint operators
$\Op{A}(y) = A(-i\hbar\partial_{\ve{x}},\ve{x},y),\,
\Op{H}^\prime(y) = H^\prime(-i\hbar\partial_{\ve{x}},\ve{x},y)$
is that they should not depend upon $\op{p}_y$ so they can be
restricted to act over small SOS-Hilbert space ${\cal L}$.
Again we define the restricted space as
${\cal M} = {\cal L}\oplus{\cal L}$ and Schr\" odinger equation
(\ref{eq:Schreq}) with (\ref{eq:hamA}) can be written as a first order
system (\ref{eq:homeq}) where (\ref{eq:applic}) should be
replaced by
$$
\op{\ma{L}}(y,E) = \pmatrix{ i\op{A}(y)/\hbar & -1\cr \op{K}^2(y,E) &
i\op{A}(y)/\hbar\cr},\quad
\op{\ma{J}} = \pmatrix{ 0 & -1\cr 1 & 0\cr}
$$
where all statements from the previous example remain valid except that
now the CSOS-CSOS propagator {\em cannot be separated to upper and lower part}
like (\ref{eq:TT}) and all blocks of $\op{\ma{T}}(E)$ are generally nonzero.

As for another interesting application one can decompose the Green function
of a relativistic Dirac $\half-$spin fermion bound in external electromagnetic
field $A^\mu(\ve{x},y)$ and search for its stationary states.
One may choose e.g. $\ve{x} = (x^1,x^2), y = x^3$ and $\op{\ma{L}}(y,E) =
\gamma^3(\gamma^0(-iE + ie A^0) + \gamma^1(\partial_1 - ie A^1) +
\gamma^2(\partial_2 - ie A^2) + im) + ie A^3,\;
\op{\ma{J}} = -\gamma^3,$ so that equations (\ref{eq:homeq},\ref{eq:inheq})
reduce to Dirac equation where $\hbar = c = 1$. The reduced Hilbert space is
now the space of Dirac spinor valued functions over 2-dim plane $(x^1,x^2)$.

\section{Multiple sections}

In this section I consider the case of multiply connected CSOS.
Let CSOS, which is now a smooth multi-sheeted $(f-1)$-dim
manifold ${\cal S}$, divide the $f-$dim CS ${\cal C}$ on countably many
disconnected parts whose closures are denoted by
${\cal C}_\alpha,\,\alpha \in {\cal J}$,
$$
{\cal C} = \bigcup\limits_{\alpha\in{\cal J}}{\cal C}_\alpha
$$
where ${\cal J}$ is some finite or countable index set.
Two points are in the same compartment ${\cal C}_\alpha$ if
they can be connected by a continuous curve which does not cross
CSOS ${\cal S}$. The compartments ${\cal C}_\alpha$ and ${\cal C}_\beta$ are
said to be neighbouring (denoted by $\alpha\vert\beta$) if their intersection
${\cal S}_{\alpha\beta}$ is nonempty $(f-1)$-dim manifold
$$
\alpha\vert\beta\quad\Leftrightarrow\quad
{\cal C}_\alpha \cap {\cal C}_\beta = {\cal S}_{\alpha\beta} \neq\emptyset.
$$
The union of all such intersections is the whole CSOS
$$
{\cal S} = \bigcup\limits_{\alpha\vert\beta} {\cal S}_{\alpha\beta}.
$$
$\Op{H}$ is self-adjoint operator over the Hilbert space
${\cal H} = L^2({\cal C})$. Let ${\cal O}_{\alpha\beta},\alpha\vert\beta$
be such open sets which cover the connected parts of CSOS,
${\cal S}_{\alpha\beta} \subset {\cal O}_{\alpha\beta}$.
Hamiltonian operator ${\cal H}$ is admissible if there exist
coordinates $(\ve{x},y)_{\alpha\beta}$ for each of the sets
${\cal O}_{\alpha\beta}$ such that
\begin{eqnarray}
\Op{H}\vert_{L^2({\cal O}_{\alpha\beta})} &=&
-\half\hbar^2 \partial_y\Op{M}^{-1}_{\alpha\beta}(y)\partial_y +
\Op{H}^\prime_{\alpha\beta}(y),\\
\Op{M}_{\alpha\beta}(y) &=& M_{\alpha\beta}(-i\hbar\partial_{\ve{x}},\ve{x},y),
\nonumber\\
\Op{H}^\prime_{\alpha\beta}(y) &=& H^\prime_{\alpha\beta}
(-i\hbar\partial_{\ve{x}},\ve{x},y).\nonumber
\end{eqnarray}
We choose the sign of coordinate $y$ of ${\cal O}_{\alpha\beta}$ so that
$(\ve{x},y)_{\alpha\beta}\in {\cal C}_{\beta}$ if $y > 0$.
Here we have allowed for very general ``masses'' $\Op{M}_{\alpha\beta}(y)$,
which should of course be positive operators and hence invertible,
which is another generalization of this section.
Then I introduce small SOS-Hilbert spaces
${\cal L}_{\alpha\beta} = L^2({\cal S}_{\alpha\beta}),\,\alpha\vert\beta$.
The operators restricted to ${\cal L}_{\alpha\beta}$ will be again denoted
by accent $\op{}$.
Now cut off the CS around ${\cal C}_\alpha$ and attach $y-$flat
the so called $\alpha\beta$-waveguides on the other sides of all connected
parts ${\cal S}_{\alpha\beta},\,\alpha\vert\beta$ of the boundary
$\partial{\cal C}_\alpha$
(see figure 2). Thus one defines the scattering Hamiltonians
which in local coordinates read
\begin{equation}
\Op{H}_\alpha\vert_{L^2({\cal O}_{\alpha\beta})} = \left\{ \begin{array}{ll}
-(\hbar^2/2)\partial_y\Op{M}^{-1}_{\alpha\beta}(y)\partial_y +
\Op{H}^\prime_{\alpha\beta}(y); & y < 0,\\
-(\hbar^2/2)\partial_y\Op{M}^{-1}_{\alpha\beta}(0)\partial_y +
\Op{H}^\prime_{\alpha\beta}(0); & y \ge 0,\end{array}
\right.
\quad \alpha\vert\beta.
\label{eq:gscH}
\end{equation}
The fundamental solution of the time-independent Schr\" odinger equation for
the scattering problem (\ref{eq:gscH}) in the $\alpha\beta$-waveguide is given
by
$$_{\beta\alpha}\bra{\ve{x}}\op{M}^{1/2}_{\alpha\beta}(0)
\op{K}^{-1/2}_{\alpha\beta}(E)e^{\pm i\op{K}_{\alpha\beta}(E) y}
$$
where the wavenumber operator $\op{K}_{\alpha\beta}(E)$ is the positive square
root of the self adjoint operator
\begin{equation}
\op{K}^2_{\alpha\beta}(E) = \frac{2}{\hbar^2} \op{M}^{1/2}_{\alpha\beta}(0)
(E - \op{H}^\prime_{\alpha\beta}(0))\op{M}^{1/2}_{\alpha\beta}(0).
\end{equation}
Vectors from the dual space ${\cal L}^\prime_{\alpha\beta}$ are
written with reversed indices, e.g. $_{\beta\alpha}\bra{\ve{x}} \in
{\cal L}^\prime_{\alpha\beta}$. Thus the general scattering wavefunction of
the Hamiltonian $\Op{H}_\alpha$ in the $\alpha\beta$-waveguide ($y > 0$) reads
\begin{eqnarray}
\Psi_\alpha(\ve{x},y,E) &=&
\frac{\sqrt{-i}}{\hbar}\ _{\beta\alpha}\bra{\ve{x}}\op{M}^{1/2}_{\alpha\beta}
\op{K}^{-1/2}_{\alpha\beta}(E)\Big[ e^{-i\op{K}_{\alpha\beta}(E)y}
\ket{\psi}_{\alpha\beta} +\nonumber\\
&+& e^{i\op{K}_{\alpha\beta}(E)y}\sum\limits_\gamma^{\gamma\vert\alpha}
\op{T}_{\beta\alpha\gamma}(E)\ket{\psi}_{\alpha\gamma}\Big],\label{eq:gWF}\\
\Psi^*_\alpha(\ve{x},y,E^*) &=&
\frac{\sqrt{-i}}{\hbar}\Big[\ _{\beta\alpha}\bra{\psi^*}
e^{-i\op{K}_{\alpha\beta}(E) y} + \nonumber \\
&+& \sum_\gamma^{\gamma\vert\alpha}\ _{\gamma\alpha}\bra{\psi^*}
\op{T}_{\gamma\alpha\beta}(E) e^{i\op{K}_{\alpha\beta}(E) y} \Big]
\op{K}^{-1/2}_{\alpha\beta}(E) \op{M}^{1/2}_{\alpha\beta}
\ket{\ve{x}}_{\alpha\beta}, \label{eq:gWFc}
\end{eqnarray}
(where $\op{M}^{1/2}_{\alpha\beta} = \op{M}^{1/2}_{\alpha\beta}(0)$)
and is uniquely determined by the incoming waves para\-me\-tri\-zed
by the SOS-states $\ket{\psi}_{\alpha\gamma}$ or
$\ket{\psi^*}_{\alpha\gamma}$ coming from the $\alpha\gamma$-waveguide
for all neighbouring compartments ${\cal C}_\gamma$.
We have introduced the scattering operators
$\op{T}_{\beta\alpha\gamma}$ which will be called generalized
CSOS-CSOS propagators. $\op{T}_{\beta\alpha\gamma}$ is the scattering
operator from ${\cal L}_{\alpha\gamma}$ to ${\cal L}_{\beta\alpha}$ and
describes the propagation from ${\cal C}_\gamma$ to ${\cal C}_\beta$
via ${\cal C}_{\alpha}$.
Then we define the two types of linear operators:
$\Oo{Q}^\prime_{\alpha\gamma}$ from small SOS-Hilbert spaces
${\cal L}_{\alpha\gamma}$ to Hilbert space ${\cal H}$, and
$\oO{P}^\prime_{\gamma\alpha}$ from Hilbert space ${\cal H}$ to
small SOS-Hilbert spaces ${\cal L}_{\gamma\alpha}$ by the following
prescriptions
\begin{eqnarray*}
\sum\limits_{\gamma\in{\cal J}}^{\gamma\vert\alpha}
\Bra{\ve{q}}\Oo{Q}^\prime_{\alpha\gamma}(E)\ket{\psi}_{\alpha\gamma} &=&
\Psi_{\alpha}(\ve{q},E), \label{eq:gQ} \\
\sum\limits_{\gamma\in{\cal J}}^{\gamma\vert\alpha}\
_{\gamma\alpha}\bra{\psi^*}
\oO{P}^\prime_{\gamma\alpha}(E)\Ket{\ve{q}} &=&
\Psi^*_{\alpha}(\ve{q},E^*) \label{eq:gP}.
\end{eqnarray*}
The resolvent of the scattering Hamiltonian with outgoing boundary
conditions
$$\Op{G}_\alpha(E) = (E - \Op{H}_\alpha + i0)^{-1}$$
can also be written explicitly  (in analogy with equation (\ref{eq:G}))
inside the waveguides ($y \ge 0, y^\prime \ge 0$)
\begin{eqnarray}
\Bra{(\ve{x},y)_{\alpha\beta}}\Op{G}_\alpha(E)
\Ket{(\ve{x}^\prime,y^\prime)_{\alpha\gamma}} &=&
\frac{1}{i\hbar^2}\ _{\beta\alpha}\bra{\ve{x}}\op{M}^{1/2}_{\alpha\beta}
\op{K}^{-1/2}_{\alpha\beta}(E)\Big[ \delta_{\beta\gamma}
e^{i\op{K}_{\alpha\beta}(E)|y - y^\prime|} + \nonumber \\
&+& e^{i\op{K}_{\alpha\beta}(E)y} \op{T}_{\beta\alpha\gamma}(E)
e^{i\op{K}_{\alpha\gamma}(E)y^\prime}\Big]
\op{K}^{-1/2}_{\alpha\gamma}(E)\op{M}^{1/2}_{\alpha\gamma}
\ket{\ve{x}^\prime}_{\alpha\gamma}.\quad\quad\label{eq:gG}
\end{eqnarray}
In analogy with the simply connected case we define also:
(i) The operators $\Oo{Q}_{\alpha\gamma}(E)$ from
${\cal L}_{\alpha\gamma}$ to ${\cal H}$ and the operators
$\oO{P}_{\gamma\alpha}(E)$ from ${\cal H}$ to
${\cal L}_{\gamma\alpha}$ with the kernels
\begin{eqnarray}
\Bra{\ve{q}}\Oo{Q}_{\alpha\gamma}(E)\ket{\psi}_{\alpha\gamma} &=&
\left\{ \begin{array}{ll}
\Bra{\ve{q}}\Oo{Q}^\prime_{\alpha\gamma}(E)\ket{\psi}_{\alpha\gamma}; &
\ve{q}\in{\cal C}_\alpha, \\
0; & \ve{q}\notin{\cal C}_\alpha, \end{array}\right. \label{eq:defgQ} \\
_{\gamma\alpha}\bra{\psi}\oO{P}_{\gamma\alpha}(E)\Ket{\ve{q}} &=&
\left\{ \begin{array}{ll}
_{\gamma\alpha}\bra{\psi}\oO{P}^\prime_{\gamma\alpha}(E)\Ket{\ve{q}};
& \ve{q}\in{\cal C}_\alpha, \\ 0;
& \ve{q}\notin{\cal C}_\alpha. \end{array}\right. \label{eq:defgP}
\end{eqnarray}
which are called generalized CSOS-CS and CS-CSOS propagators respectively,
and (ii) $\Op{G}_0(E)$ a linear operator over ${\cal H}$ with the kernel
\begin{equation}
\Bra{\ve{q}}\op{G}_0(E)\Ket{\ve{q}^\prime} =
\left\{ \begin{array}{ll}
\Bra{\ve{q}}\op{G}_\alpha(E)\Ket{\ve{q}^\prime};
& \exists\alpha,\; \ve{q},\ve{q}^\prime \in{\cal C}_\alpha,\\
0; & {\rm otherwise.} \end{array} \right.
\label{eq:gG0}
\end{equation}
which is called generalized CS-CS propagator (without crossing the CSOS
in between).

Let us compactify our notation by introducing the following symbols.
First we define the large SOS-Hilbert space ${\cal M}$
$$
{\cal M} = \bigoplus\limits_{\alpha\vert\beta} {\cal L}_{\alpha\beta}
$$
with a complete system of orthogonal projectors $\op{\Pi}_{\alpha\beta}$.
(Note that each pair $(\alpha,\beta)$ is always included twice, once
as $\alpha\vert\beta$ and once as $\beta\vert\alpha$.)
For each SOS-state $\ket{\psi}$ we write symbolically
$$
\ket{\psi} = \sum\limits_{\alpha\vert\beta}\ket{\psi}_{\alpha\beta},
\quad \ket{\psi}_{\alpha\beta} = \op{\Pi}_{\alpha\beta}\ket{\psi}.
$$
One then defines the large operators $\op{T}(E)$, $\Oo{Q}(E)$ and $\oO{P}(E)$
by
\begin{eqnarray*}
\op{T}(E) &=& \sum\limits_{\beta\vert\alpha\vert\gamma}
\op{T}_{\beta\alpha\gamma}(E)\op{\Pi}_{\alpha\gamma},\\
\Oo{Q}(E) &=& \sum\limits_{\alpha\gamma}
\Oo{Q}_{\alpha\gamma}(E)\op{\Pi}_{\alpha\gamma},\\
\oO{P}(E) &=& \sum\limits_{\gamma\alpha}
\op{\Pi}_{\gamma\alpha}\oO{P}_{\gamma\alpha}(E),
\end{eqnarray*}
or equivalently
\begin{eqnarray}
\op{\Pi}_{\beta\alpha}\op{T}(E) &=&
\sum_{\gamma\in{\cal J}}^{\gamma\vert\alpha}
\op{T}_{\beta\alpha\gamma}(E)\op{\Pi}_{\alpha\gamma},\label{eq:lT}\\
\Oo{Q}(E)\op{\Pi}_{\alpha\gamma} &=&
\Oo{Q}_{\alpha\gamma}(E)\op{\Pi}_{\alpha\gamma},\label{eq:lQ}\\
\op{\Pi}_{\gamma\alpha}\oO{P}(E) &=& \op{\Pi}_{\gamma\alpha}
\oO{P}_{\gamma\alpha}(E).
\label{eq:lP}
\end{eqnarray}
Now, geometrically most general form of the main result of this paper can be
stated and proved very elegantly.
\\\\
{\bf Theorem 2c:} The resolvent of the Hamiltonian $\Op{G}(E) =
(E-\Op{H})^{-1}$
can be decomposed in terms of four elementary propagators, namely
$\Op{G}_0 : {\cal H}\rightarrow {\cal H}$,
$\Oo{Q} : {\cal M}\rightarrow {\cal H}$,
$\oO{Q} : {\cal H}\rightarrow {\cal M}$, and
$\op{T} : {\cal M}\rightarrow {\cal M}$, as follows
\begin{equation}
\Op{G}(E) = \Op{G}_0(E) + \Oo{Q}(E) \left(1 - \op{T}(E)\right)^{-1}\oO{P}(E).
\label{eq:gdecomp}
\end{equation}
\\\\
{\bf Proof:} Put the decomposition formula in a sandwitch between
$\Bra{\ve{q}}$ and $\Ket{\ve{q}^\prime}$. One should prove that RHS also
solves the inhomogeneous Schr\" odinger equation as the LHS does
\begin{equation}
(E - H(-i\hbar\partial_{\ve{q}},\ve{q}))\Bra{\ve{q}}\Op{G}(E)
\Ket{\ve{q}^\prime} = \delta(\ve{q}-\ve{q}^\prime).
\label{eq:Schr}
\end{equation}
This is indeed true in every compartment ${\cal C}_\alpha$ separately, where
$\Bra{\ve{q}}\Op{G}_0(E)\Ket{\ve{q}^\prime}$ is a particular solution and
$\Bra{\ve{q}}\Oo{Q}(E)(1 - \op{T}(E))^{-1}\oO{P}(E)\Ket{\ve{q}^\prime}$
is a solution of the homogeneous equation by construction of operator
$\Oo{Q}(E)$. What is left to prove is that RHS is continuously
differentiable on borders between compartments, that is on CSOS ${\cal S}$.
Take arbitrary neighbouring compartments ${\cal C}_\alpha$ and
${\cal C}_\beta$ and
choose coordinates $(\ve{x},y)_{\alpha\beta}$ of an open set
${\cal O}_{\alpha\beta}$ which includes ${\cal S}_{\alpha\beta}$.
We shall need the following values and normal derivatives of the
CSOS-CS and CS-CSOS propagators on the ${\cal S}_{\alpha\beta}$ which can
be obtained directly from (\ref{eq:gWF},\ref{eq:gWFc},\ref{eq:gQ},
\ref{eq:gP})
\begin{eqnarray}
\Bra{(\ve{x},0)_{\alpha\beta}}\Oo{Q}_{\alpha\gamma}(E) &=&
\frac{\sqrt{-i}}{\hbar}\ _{\beta\alpha}\bra{\ve{x}}\op{M}^{1/2}_{\alpha\beta}
\op{K}^{-1/2}_{\alpha\beta}(E)\left(
\op{T}_{\beta\alpha\gamma}(E) + \delta_{\beta\gamma}\right), \label{eq:vQ}\\
\partial_y\Bra{(\ve{x},0)_{\alpha\beta}}\Oo{Q}_{\alpha\gamma}(E)\vert_{y=0}&=&
\frac{\sqrt{i}}{\hbar}\ _{\beta\alpha}\bra{\ve{x}}\op{M}^{1/2}_{\alpha\beta}
\op{K}^{1/2}_{\alpha\beta}(E)\left(
\op{T}_{\beta\alpha\gamma}(E) - \delta_{\beta\gamma}\right), \label{eq:dQ}\\
\oO{P}_{\gamma\alpha}(E)\Ket{(\ve{x},0)_{\alpha\beta}} &=&
\frac{\sqrt{-i}}{\hbar}\left(\op{T}_{\gamma\alpha\beta}(E) +
\delta_{\beta\gamma}\right)\op{K}^{-1/2}_{\alpha\beta}(E)
\op{M}^{1/2}_{\alpha\beta}\ket{\ve{x}}_{\alpha\beta}, \label{eq:vP}\\
\partial_y\oO{P}_{\gamma\alpha}(E)\Ket{(\ve{x},y)_{\alpha\beta}}\vert_{y=0} &=&
\frac{\sqrt{i}}{\hbar}\left(\op{T}_{\gamma\alpha\beta}(E) -
\delta_{\beta\gamma}\right)\op{K}^{1/2}_{\alpha\beta}(E)
\op{M}^{1/2}_{\alpha\beta}\ket{\ve{x}}_{\alpha\beta}. \label{eq:dP}
\end{eqnarray}

First we shall prove that RHS of (\ref{eq:gdecomp}) is continuous
on all ${\cal S}_{\alpha\beta},\, \alpha\vert\beta$.
Using the difference operator (\ref{eq:diffop})
we can write (omitting the argument $(E)$ for the sake of brevity)
\begin{eqnarray*}
&&\Delta_y \Bra{(\ve{x},0)_{\alpha\beta}}\Op{G}\Ket{\ve{q}^\prime} -
\Delta_y \Bra{(\ve{x},0)_{\alpha\beta}}\Op{G}_0\Ket{\ve{q}^\prime}=\\
&&=\Bra{(\ve{x},0)_{\alpha\beta}}
\left(\sum_{\gamma\in{\cal J}}^{\gamma\vert\alpha}
\Oo{Q}_{\alpha\gamma}\op{\Pi}_{\alpha\gamma}
- \sum_{\gamma\in{\cal J}}^{\gamma\vert\beta}
\Oo{Q}_{\beta\gamma}\op{\Pi}_{\beta\gamma}\right)
(1-\op{T})^{-1}\oO{P}\Ket{\ve{q}^\prime}=\\
&&=\frac{\sqrt{-i}}{\hbar}\ _{\beta\alpha}\bra{\ve{x}}
\op{M}^{1/2}_{\alpha\beta}\op{K}^{-1/2}_{\alpha\beta}
\left[\sum_{\gamma\in{\cal J}}^{\gamma\vert\alpha}
(\op{T}_{\beta\alpha\gamma} + \delta_{\beta\gamma})\op{\Pi}_{\alpha\gamma} -
\sum_{\gamma\in{\cal J}}^{\gamma\vert\beta}
(\op{T}_{\alpha\beta\gamma} + \delta_{\alpha\gamma})\op{\Pi}_{\beta\gamma}
\right](1-\op{T})^{-1}\oO{P}\Ket{\ve{q}^\prime}=\\
&&=\frac{\sqrt{-i}}{\hbar}\ _{\beta\alpha}\bra{\ve{x}}
\op{M}^{1/2}_{\alpha\beta}\op{K}^{-1/2}_{\alpha\beta}
\left[\oO{P}_{\alpha\beta}-\oO{P}_{\beta\alpha}\right]\Ket{\ve{q}^\prime}.
\end{eqnarray*}
We have applied equations (\ref{eq:vQ},\ref{eq:lT},\ref{eq:lP}). And
analogously, by applying equations (\ref{eq:dQ},\ref{eq:lT},\ref{eq:lP}) we get
(note also that $(\ve{x},y)_{\alpha\beta} = (\ve{x},-y)_{\beta\alpha}$)
\begin{eqnarray*}
&&\Delta_y \partial_y\Bra{(\ve{x},y)_{\alpha\beta}}\Op{G}\Ket{\ve{q}^\prime}
\vert_{y=0} -
\Delta_y \partial_y\Bra{(\ve{x},y)_{\alpha\beta}}\Op{G}_0\Ket{\ve{q}^\prime}
\vert_{y=0} =\\
&&=\frac{\sqrt{i}}{\hbar}\ _{\beta\alpha}\bra{\ve{x}}
\op{M}^{1/2}_{\alpha\beta}\op{K}^{1/2}_{\alpha\beta}
\left[\sum_{\gamma\in{\cal J}}^{\gamma\vert\alpha}
(\op{T}_{\beta\alpha\gamma} - \delta_{\beta\gamma})\op{\Pi}_{\alpha\gamma} +
\sum_{\gamma\in{\cal J}}^{\gamma\vert\beta}
(\op{T}_{\alpha\beta\gamma} - \delta_{\alpha\gamma})\op{\Pi}_{\beta\gamma}
\right](1-\op{T})^{-1}\oO{P}\Ket{\ve{q}^\prime}=\\
&&=\frac{\sqrt{i}}{\hbar}\ _{\beta\alpha}\bra{\ve{x}}
\op{M}^{1/2}_{\alpha\beta}\op{K}^{1/2}_{\alpha\beta}
\left[\oO{P}_{\alpha\beta}+\oO{P}_{\beta\alpha}\right]\Ket{\ve{q}^\prime}.
\end{eqnarray*}
In order to see that
$\Delta_y \Bra{(\ve{x},0)_{\alpha\beta}}\Op{G}\Ket{\ve{q}^\prime} = 0$ and
$\Delta_y \partial_y\Bra{(\ve{x},y)_{\alpha\beta}}\Op{G}\Ket{\ve{q}^\prime}
\vert_{y=0} = 0$ one has to prove
\begin{eqnarray}
\Bra{(\ve{x},0)_{\alpha\beta}}\Op{G}_\alpha(E)\Ket{\ve{q}^\prime} &=&
\frac{\sqrt{-i}}{\hbar}\ _{\beta\alpha}\bra{\ve{x}}\op{M}^{1/2}_{\alpha\beta}
\op{K}^{-1/2}_{\alpha\beta}(E)\oO{P}_{\beta\alpha}(E)\Ket{\ve{q}^\prime},
\label{eq:bbb}\\
\partial_y\Bra{(\ve{x},y)_{\alpha\beta}}\Op{G}_\alpha(E)\Ket{\ve{q}^\prime}
\vert_{y=0} &=&
\frac{\sqrt{i}}{\hbar}\ _{\beta\alpha}\bra{\ve{x}}\op{M}^{1/2}_{\alpha\beta}
\op{K}^{1/2}_{\alpha\beta}(E)\oO{P}_{\beta\alpha}(E)\Ket{\ve{q}^\prime},
\label{eq:ccc}
\end{eqnarray}
considering the definition of $\Op{G}_0(E)$ in terms of $\Op{G}_\alpha(E)$
(\ref{eq:gG0}).
But this is easy. Both, LHSs and RHSs of (\ref{eq:bbb},\ref{eq:ccc}) satisfy
the conjugated Schr\" odinger equation as functions
of $\ve{q}^\prime$. The initial data --- the values and the
normal derivatives of LHSs and RHSs on any initial surface
$(\ve{x},y)_{\alpha\gamma},\,\alpha\vert\gamma$ also match as can be
seen by applying $\ve{q}^\prime = (\ve{x},y)_{\alpha\gamma}$ and
(\ref{eq:gG}) to LHSs and (\ref{eq:vP},\ref{eq:dP}) to RHSs.
Formulas (\ref{eq:bbb},\ref{eq:ccc})/(\ref{eq:gdecomp}) then follow from the
uniqueness of the initial value homoge\-ne\-us
(\ref{eq:Schreq})/non-homogeneous (\ref{eq:Schr}) Schr\" odinger problem.
q.e.d.

One can formally expand decomposition formula (\ref{eq:gdecomp}) in a
geometric series or {\em sum over paths}
\begin{equation}
\Op{G}(E) = \Op{G}_0(E) +
\sum_{\alpha_1\vert\alpha_2\ldots\vert\alpha_n}^{n\ge 2}
\Oo{Q}_{\alpha_n \alpha_{n-1}}(E)\op{T}_{\alpha_n\alpha_{n-1}\alpha_{n-2}}(E)
\ldots \op{T}_{\alpha_3\alpha_2\alpha_1}(E)\oO{P}_{\alpha_2\alpha_1}(E)
\quad\quad\label{eq:sumoverpaths}
\end{equation}
where each term contains probability amplitudes to propagate from
compartment ${\cal C}_{\alpha_1}$ to ${\cal C}_{\alpha_2} \ldots$
to ${\cal C}_{\alpha_N}$. If one chooses many disconnected parts
of CSOS ${\cal S}_{\alpha_k \alpha_{k+1}}$ which are geometrically close
then the propagators $\op{T}_{\alpha\beta\gamma}(E)$ would become simple
and they could be asymptotically explicitly calculated, so the formula
(\ref{eq:sumoverpaths}) would turn into a kind of path integral formula
for energy dependent quantum propagator.

\section{Discussion and conclusions}

This paper presents theoretical construction of SOS
reduction of quantum dynamics in analogy with the SOS reduction of
classical dynamics. However, there is an important difference:
In classical dynamics, one should carefully choose SOS such that
almost every trajectory crosses it, while in quantum dynamics this is
not essential. All theorems work even if CSOS lies in classically forbidden
region although the practical usefulness of the method is expected to be worse
then, because of the exponential localization and sensitive dependence on
boundary conditions. Moreover, formalism of section 4 can be
easily adapted (by taking two different CSOSs as a single multiply connected
CSOS) to show explicitly that the spectra as determined by our method do not
depend on the choice of the CSOS, since the corresponding quantum Poincar\' e
mappings are related to each other by a kind of similarity transformation.

The Green function -- energy dependent quantum propagator has been
decomposed in terms of propagators which propagate from CS/CSOS to
CS/CSOS. This decomposition formula has been generalized in two ways:
(i) for Green functions of arbitrary linear differential systems
and (ii) for SOS which consists of more than one disconnected part.
The combination of these two generalizations is straigtforward so it is
not given explicitly in this paper. While this general decomposition formula
((\ref{eq:gdecomp}) or even (\ref{eq:adecomp})) has so far merely theoretical
value, it has a very practical consequence, namely, SOS-quantization condition:
The resolvent of the Hamiltonian $(E-\Op{H})^{-1}$ can have a pole,
i.e. eigenenergy $E_0$, only when the operator $1 - \op{T}(E_0)$ is singular,
i.e. when {\em general quantum Poincar\' e mapping} $\op{T}(E_0)$ has a fixed
point $\ket{\psi}\in{\cal M}$, $\op{T}(E_0)\ket{\psi} = \ket{\psi}$. For the
more special and common case of section 2 we have
${\cal M} = {\cal L}\oplus{\cal L}$,
$\op{T} = \pmatrix{0 & T_\uparrow\cr T_\downarrow & 0\cr}$, and
$\ket{\psi} = \pmatrix{\ket{\uparrow} \cr \ket{\downarrow}\cr}$, where
$\ket{\uparrow}$ is a fixed point of quantum Poincar\' e mapping
$\op{T}_\downarrow\op{T}_\uparrow$ and at the same time $\ket{\downarrow}$ is
a fixed point of a similar mapping $\op{T}_\uparrow\op{T}_\downarrow$.
This quantization condition can be very efficiently numerically
implemented \cite{SS93,RS94,P94c}. Since the exact
quantum Poincar\' e mapping is usually difficult to calculate explicitly we
describe its semiclassical $\hbar-$expansion and give explicitly the
leading (Bogomolny's \cite{B92}) and next-to-leading order terms.

\section*{Acknowledgments}

I am grateful to Professor Marko Robnik for fruitful discussions.
The financial support by the Ministry of Science
and Technology of the Republic of Slovenia is gratefully acknowledged.

\vfill
\newpage
\section*{Figures}
\bigskip
\bigskip

\noindent{\bf Figure 1} The geometry of 2-dim CS
of a typical bound system (a) with simply connected CSOS.
Isopotential contours are shown. The product of classical or quantal
scattering mappings of the two scattering systems shown in (b,c) is equal to
the classical or quantal Poincar\' e mapping of a bound system (a).

\bigskip
\bigskip
\noindent{\bf Figure 2} The geometry of 2-dim CS of a bound system
with multiple sections (a). One of the related scattering systems is
shown schematically in (b).
\end{document}